\documentclass[aps,prc,preprint,superscriptaddress,showpacs]{revtex4-1}

\usepackage{graphicx}% Include figure files
\usepackage{dcolumn}% Align table columns on decimal point
\usepackage{longtable}
\usepackage{placeins}
\usepackage[dvipdfm]{hyperref}
\usepackage{multirow}
\usepackage{lineno}

\hypersetup{ 
    colorlinks=true, 
    citecolor=black, 
    filecolor=black, 
    linkcolor=black, 
    urlcolor=black 
} 

\begin{document}

%\preprint{}

\title{Measurement and analysis of the \texorpdfstring{$^{243}$Am}{Am-243} neutron capture cross section at the n\_TOF facility at CERN}

\author{E.~Mendoza} \affiliation{CIEMAT, Centro de Investigaciones Energ\'{e}ticas Medioambientales y Tecn\'{o}logicas, Madrid, Spain}
\author{D.~Cano-Ott} \affiliation{CIEMAT, Centro de Investigaciones Energ\'{e}ticas Medioambientales y Tecn\'{o}logicas, Madrid, Spain}
\author{C.~Guerrero} \affiliation{CIEMAT, Centro de Investigaciones Energ\'{e}ticas Medioambientales y Tecn\'{o}logicas, Madrid, Spain} \affiliation{CERN, Geneva, Switzerland}
\author{E.~Berthoumieux} \affiliation{CEA/Saclay - IRFU, Gif-sur-Yvette, France}
\author{U.~Abbondanno} \affiliation{Istituto Nazionale di Fisica Nucleare, Trieste, Italy}
\author{G.~Aerts} \affiliation{CEA/Saclay - IRFU, Gif-sur-Yvette, France}
\author{F.~\'{A}lvarez-Velarde} \affiliation{CIEMAT, Centro de Investigaciones Energ\'{e}ticas Medioambientales y Tecn\'{o}logicas, Madrid, Spain}
\author{S.~Andriamonje} \affiliation{CEA/Saclay - IRFU, Gif-sur-Yvette, France}
\author{J.~Andrzejewski} \affiliation{University of Lodz, Lodz, Poland}
\author{P.~Assimakopoulos} \affiliation{University of Ioannina, Greece}
% ...
 
%\collaboration{The n\_TOF Collaboration $\langle$www.cern.ch/nTOF$\rangle$}  \noaffiliation
\collaboration{The n\_TOF Collaboration}  \noaffiliation

\date{\today}

\begin{abstract}
\begin{description}
\item[Background] The design of new nuclear reactors and transmutation devices requires to reduce the present neutron cross section uncertainties of minor actinides.
\item[Purpose] Reduce the $^{243}$Am(n,$\gamma$) cross section uncertainty.
\item[Method] The $^{243}$Am(n,$\gamma$) cross section has been measured at the n\_TOF facility at CERN with a BaF$_{2}$ Total Absorption Calorimeter, in the energy range between 0.7 eV and 2.5 keV.
\item[Results] The $^{243}$Am(n,$\gamma$) cross section has been successfully measured in the mentioned energy range. The resolved resonance region has been extended from 250 eV up to 400 eV. In the unresolved resonance region our results are compatible with one of the two incompatible capture data sets available below 2.5 keV. The data available in EXFOR and in the literature has been used to perform a simple analysis above 2.5 keV.
\item[Conclusions] The results of this measurement contribute to reduce the $^{243}$Am(n,$\gamma$) cross section uncertainty and suggest that this cross section is underestimated up to 25\% in the neutron energy range between 50 eV and a few keV in the present evaluated data libraries.
\end{description}
\end{abstract}

%\keywords{}

\maketitle

% =========================================================================================================
% =========================================================================================================
%                                             INTRODUCTION
% =========================================================================================================

\section{Introduction}{\label{sec:Introduction}}
\FloatBarrier

Nuclear data for minor actinides have gained importance in the last years because they are necessary for improving the design and performance of advanced nuclear reactors and transmutation devices for the incineration of radioactive nuclear waste~\cite{NDATA01,NDATA02,NDATA03}. In particular, $^{243}$Am is the minor actinide which contributes most to the total radiotoxicity of the spent fuel at times after disposal close to its half life (7370 years). In addition, in a nuclear reactor most of the production of $^{244}$Cm, which is a strong neutron emitter and which is in the path of the creation of any heavier isotope, is originated as the result of the $^{243}$Am(n,$\gamma$) reaction.

\begin{table}[bt]
\caption{\label{tab:differential_measurements}
Differential measurements performed up to now relevant for the evaluation of the $^{243}$Am capture cross section.
}
\begin{ruledtabular}
\begin{tabular}{lcc}
Reference & Type & Range (eV) \\
\colrule
Bellanova \textit{et al.} (1976)~\cite{Bellanova} & Transmission & 0.35 -- 35 \\
Simpson \textit{et al.} (1974)~\cite{Simpson} & Transmission & 0.5 -- 1$\cdot$10$^{3}$ \\
Berreth \textit{et al.} (1970)~\cite{Berreth} & Transmission & 0.008 -- 25.6 \\
Cote \textit{et al.} (1959)~\cite{Cote} & Transmission & 0.0014 -- 15.44 \\
Weston \textit{et al.} (1985)~\cite{Weston} & Capture & 258 -- 9.2$\cdot$10$^{4}$ \\ 
Wisshak \textit{et al.} (1983)~\cite{Wisshak} & Capture & 5$\cdot$10$^{3}$ -- 2.5$\cdot$10$^{5}$ \\
\colrule
Jandel \textit{et al.} (2009)~\cite{Jandel} \footnotemark[1] & Capture & 8 -- 2.5$\cdot$10$^{5}$ \\
Hori \textit{et al.} (2009)~\cite{Hori} \footnotemark[1] & Capture & 0.01 -- 400 \\
This work & Capture & 0.7 -- 2.5$\cdot$10$^{3}$ \\
\colrule
Kimura \textit{et al.} (2012)~\cite{Kimura} \footnotemark[2] & Capture & -- \\
Alekseev \textit{et al.} (2011)~\cite{Alekseev} \footnotemark[2] & Fission &  -- \\
\end{tabular}
\end{ruledtabular}
\footnotetext[1]{Neither the yield nor the resulting cross sections have been published yet.}
\footnotetext[2]{Only the resonance parameters of the resonance at 1.35 eV (Kimura \textit{et al.}) or below 17 eV (Alekseev \textit{et al.}) are provided.}
\end{table}

The differential data available for the evaluation of the $^{243}$Am capture cross section are presented in Table~\ref{tab:differential_measurements}. As it can be observed, there are only two differential capture measurements covering the energy region below 250 eV, apart from the one presented here. Both of them are recent and their final results have not been published yet. In this energy range, only the information provided by the transmission measurements have been used to determine the $^{243}$Am capture cross section in the current evaluated data libraries (the last releases at this moment are ENDF/B-VII.1~\cite{ENDF/B-VII.1}, JENDL-4.0~\cite{JENDL-4.0}, JEFF-3.1.2~\cite{JEFF-3.1.2}, ROSFOND-2010~\cite{ROSFOND-2010} and CENDL-3.1~\cite{CENDL-3.1}). In particular, the present evaluations are based essentially in the Simpson \textit{et al.} results, which are the only ones which extend above 35 eV. This information has been completed with the integral measurements presented in Table~\ref{tab:integral_measurements}, which provide the thermal capture cross section and resonance integral measurements performed up to now. As it can be observed, there are sizeable differences between them.

At higher neutron energies there are only two data sets between 250 eV and 5 keV, both of them carried out by Weston \textit{et al.}, which differ significantly below 2 keV. In addition, the results of Wisshak \textit{et al.} are 10-15\% lower than the Weston \textit{et al.} data in the energy range of overlap. Together with these differential measurements, there are also integral measurements carried out in fast nuclear reactors, which provide information of the $^{243}$Am capture cross section in the fast energy range. The results of the calculations performed with the evaluated libraries do not reproduce necessarily these experimental results~\cite{Palmiotti_integral,ENDF/B-VII.1_benchmark}. These inconsistencies have motivated, for example, changes in the evaluated $^{243}$Am capture cross section in the ENDF/B-VII.1 release with respect to ENDF/B-VII.0~\cite{ENDF/B-VII.1}.

\begin{table}[htb]
\caption{\label{tab:integral_measurements}
Thermal capture cross sections ($\sigma_{0}$), resonance integrals ($I_{0}=\int_{0.5eV}^{\infty}\sigma_{\gamma}(E)/EdE$) and ratios between them provided by different experiments and evaluations.
}
\begin{ruledtabular}
\begin{tabular}{lccc}
Reference & $\sigma_{0}$(barn) & $I_{0}$ (barn) & $I_{0}/\sigma_{0}$\\
\colrule
Hori \textit{et al.} (2009)~\cite{Hori} & 76.6\footnotemark[1] & 1970(110) & 25.7(15) \\
Marie \textit{et al.} (2006)~\cite{Marie} & 81.8(36) &  & \\
Ohta \textit{et al.} (2006)~\cite{Ohta} &  & 2250(300)\footnotemark[2] & \\
Hatsukawa \textit{et al.} (1997)~\cite{Hatsukawa} & 84.4 & & \\
Gavrilov \textit{et al.} (1977)~\cite{Gavrilov} & 83(6) & 2200(150) & 26.5(26) \\ 
Simpson \textit{et al.} (1974)~\cite{Simpson} &  & 1819(80)\footnotemark[3] & \\
Eberle \textit{et al.} (1971)~\cite{Eberle} & 77(2) & 1930(50)\footnotemark[3] & 25.1(9) \\
Berreth \textit{et al.} (1970)~\cite{Berreth} & 85(4) & 1824(80)\footnotemark[3] & 21.5(14) \\
Folger \textit{et al.} (1968)~\cite{Folger} & 78 & 2250\footnotemark[4] & 29 \\
Bak \textit{et al.} (1967)~\cite{Bak} & 73(6) &  2300(200) & 32(4) \\
Ice (1966)~\cite{Ice} & 66-84 &  & \\
Butler \textit{et al.} (1957)~\cite{Butler} & 73.6(1.8) & 2290(50) & 31(1) \\
Harvey \textit{et al.} (1954)~\cite{Harvey} & 140(50) &  & \\
Stevens \textit{et al.} (1954)~\cite{Stevens} & 115 &  & \\
\colrule
Mughabghab (2006)~\cite{Mughabghab_atlas2006} & 75.1(18) &  1820(70) & 24.2(11) \\
ENDF/B-VII.1~\cite{ENDF/B-VII.1} & 80.4 & 2051 & 25.5 \\
ENDF/B-VII.0~\cite{ENDF/B-VII.0} & 75.1 & 1820 & 24.2 \\
JENDL-4.0~\cite{JENDL-4.0} & 79.3 & 2040 & 25.7 \\
JEFF-3.1~\cite{JEFF-3.1} & 76.7 & 1788 & 23.3 \\
\end{tabular}
\end{ruledtabular}
\footnotetext[1]{Value assumed for normalization. $I_{0}$ is proportional to it.}
\footnotetext[2]{The thermal value of Marie \textit{et al.} was assumed. The Ohta \textit{et al.} measured value was $\hat{\sigma}=$174.5(5.3) barn and $\alpha$=0.0418(45), where $I_{0}=\hat{\sigma}/\alpha+(0.45-1/\alpha)\sigma_{0}$.}
\footnotetext[3]{Cut-off energy was taken as 0.625 instead of 0.5 eV.}
\footnotetext[4]{Cut-off energy was taken as 0.83 instead of 0.5 eV.}
\end{table}

The lack of data, the inconsistencies presented above, and the recent interest in the design of new nuclear devices, specially those related with the transmutation of the spent fuel, have motivated new $^{243}$Am capture cross section measurements, such as the one presented in this work or the ones of Jandel \textit{et al.} and Hori \textit{et al.}.

The experimental setup of the $^{243}$Am(n,$\gamma$) measurement carried out at the n\_TOF facility at CERN is described in Section~\ref{sec:Measurement}. The reduction of the data which leads to the capture yield, which will be available in the EXFOR database~\cite{EXFOR}, is presented in Section~\ref{sec:DataReduction}; and the cross section analysis performed with the resulting yield, in Section~\ref{sec:Analysis}. At the end of Section~\ref{sec:Analysis} we extend the analysis of the $^{243}$Am(n,$\gamma$) cross section up to higher energies with the data availble in EXFOR and in the literature. Finally, the conclusions of this work are presented in Section~\ref{sec:Conclusions}.

\FloatBarrier
% =========================================================================================================

% =========================================================================================================
% =========================================================================================================
%                                             EXP. SETUP
% =========================================================================================================

\section{The experimental setup}{\label{sec:Measurement}}

\subsection{The n\_TOF facility at CERN}{\label{sec:theFacility}}

The n\_TOF (Phase-1\footnote{The n\_TOF facility was closed at the end of the 2004 campaign. It was opened again in 2009 (Phase-2), with a different lead block and coolant circuit.}) facility at CERN ~\cite{nTOF} is a pulsed neutron source coupled to a 200 m flight path designed to study neutron-nucleus interactions for neutron kinetic energies ranging from a few meV to several GeV. The neutrons are produced in spallation reactions induced by a 20 GeV/c proton beam with 16 ns FWHM time resolution and a repetition rate of $\sim$0.4 Hz. The spallation source was a 80x80x60 cm$^{3}$ lead block surrounded by 5.8 cm of water, serving as a coolant and as a moderator for the initially fast neutron spectrum. The neutrons travel along a beam line in vacuum orientated at 10$^{\circ}$ with respect to the proton beam until reaching the measuring station. Along the beam line a magnet avoids the charged particles reaching the measuring station and two collimators give the appropriate shape to the neutron beam. This facility is used mainly to measure fission and capture cross sections relevant for nuclear astrophysics and nuclear technologies.

There are around 1.54$\cdot$10$^{5}$ neutrons per nominal pulse of 7$\cdot$10$^{7}$ protons between 1 eV and 10 keV reaching the irradiation position, placed at 185 m from the spallation source, with a nearly isolethargic energy distribution. Only proton pulses with intensities close to the mentioned nominal intensity have been considered in this analysis. At the irradiation position the neutron beam has a spatial distribution which does not vary significantly in the energy range of this measurement and that resembles a 2D-Gaussian with $\sigma_{x}$=$\sigma_{y}$=0.54 cm~\cite{nTOFBeamProfile}. The description of the resolution function can be found in~\cite{SAMMY}.

\subsection{The detection system}

Three different detectors were used to monitor the beam during the $^{243}$Am(n,$\gamma$) measurement: a wall current monitor~\cite{nTOF} and wall current transformers ~\cite{nTOF}, used to monitor the intensity of the proton beam; and a silicon flux monitor~\cite{Marrone_SiMon} used to monitor the intensity of the neutron beam. The latter is a $^{6}$Li-based silicon monitor placed around 2 m before the irradiated sample.

\begin{figure}[htb]
\begin{center}
\includegraphics[width=8.6cm]{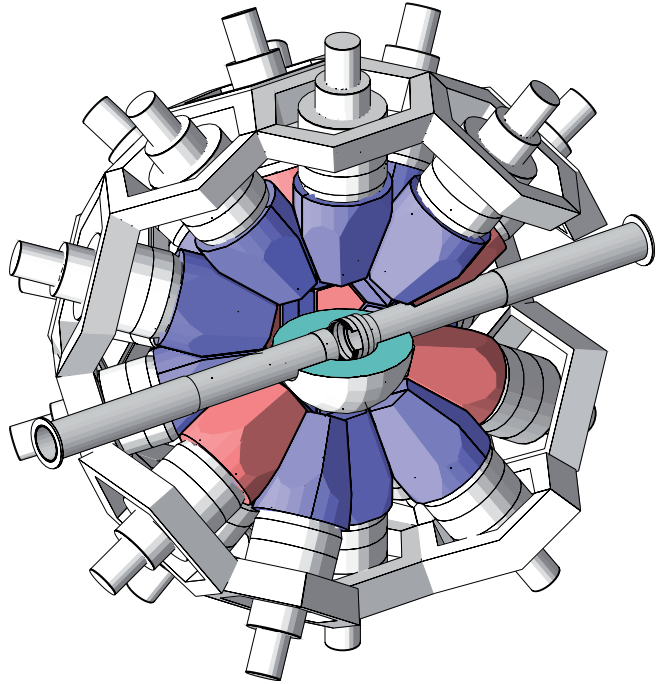} 
\caption{(Color online) Schematic view of the n\_TOF Total Absorption Calorimeter.}
\label{fig:TAC_geom}
\end{center}
\end{figure}

The $^{243}$Am(n,$\gamma$) reactions were measured with the n\_TOF Total Absorption Calorimeter (TAC)~\cite{Guerrero_TAC}, by measuring in coincidence the $\gamma$-cascades which follows the neutron capture reactions. The TAC, shown in Figure~\ref{fig:TAC_geom}, is a 4$\pi$ segmented array made of 40 BaF$_{2}$ crystals with pentagonal and hexagonal shapes. Each crystal has been constructed by cutting a BaF$_{2}$ cylinder of 14 cm diameter and 15 cm length. For optimal light collection each crystal is covered with two layers of 0.1 mm thick Teflon foil and a 0.1 mm thick polished aluminum sheet on the outside. In order to minimize the detection of scattered neutrons, the crystals are encapsulated inside 1 mm thick $^{10}$B loaded carbon fiber capsules. Each capsule is coupled to an aluminum cylinder that houses a 12.7 cm Photonis XP4508B photomultiplier and a special voltage divider made at the Instituto Tecnol\'{o}gico e Nuclear in Lisbon that favors its fast recovery. The complete modules are attached to an aluminum honey comb structure that holds the complete assembly. The TAC is divided in two hemispheres that can be opened and closed, and form a spherical shell of 10 cm inner radius and 25 cm outer radius, approximately, covering around 95\% of the entire solid angle. A neutron absorber which consists on a 5 cm thick spherical shell made of Li$_{2}$C$_{12}$H$_{20}$O$_{4}$ was placed in the inner hole of the TAC, in order to reduce, together with the $^{10}$B loaded carbon fiber capsules, the detection of neutrons scattered in the center of the TAC, where the $^{243}$Am sample was placed.

The detector signals were recorded by a digital data acquisition system~\cite{nTOF_DAQ} which used Acqiris-DC270 digitizers with 8 bits resolution operating at 250 MHz and recording continuously a time of flight of 16 ms for each pulse, thus containing the digitized electronic response of each detector for neutron energies above 0.7 eV. The data buffers were then analyzed offline, with dedicated pulse shape reconstruction algorithms. The algorithm used to analyze the BaF$_{2}$ signals is described in~\cite{TAC_PuseShape}, and a more accessible reference of a similar routine is~\cite{TAC_PuseShape2}. It returns for each signal the time-of-flight, the area, and other parameters used to distinguish the detected particle type: $\gamma$ or $\alpha$ (the latter is produced by the decay of Ra impurities in the crystals). Each detector was calibrated in energy from measurements performed with standard calibration sources ($^{137}$Cs, $^{60}$Co, $^{88}$Y, $^{24}$Na, and Pu/C), and the gain drifts occurred along the entire measurement were monitored with the changes observed in the $\alpha$ deposited energy spectra in each BaF$_{2}$ detector. The individual detector signals are grouped into TAC events using a coincidence window of 20 ns. Each TAC event is characterized by its time-of-flight, total energy deposited (E$_{Sum}$) and crystal multiplicity (m$_{cr}$), which is the number of detectors contributing to the event. The E$_{Sum}$ and m$_{cr}$ values are used to apply conditions to the detected events in order to improve the capture over background ratio. In this paper, the word \emph{event} always refers to these TAC events.

\subsection{The \texorpdfstring{$^{243}$Am}{Am-243} and auxiliary samples and measurements}{\label{sec:theSample}}

The $^{243}$Am sample was manufactured at IPPE Obninsk (Russia) in February 2004. It was in form of oxide power (AmO$_{2}$) deposited on an Al backing of 10 mm diameter and less than 70 mg, according to the specifications provided by the manufacturers. The sample was encapsulated inside a Ti canning of 15 mm diameter and 0.17 and 0.18 mm thickness above and below the sample. The whole sample (AmO$_{2}$, Al backing and Ti canning) was weighted at CERN, obtaining a value of 420.9(1) mg. According to the specifications provided by the manufacturers, the total mass of the AmO$_{2}$ deposit was 11.3 mg, and the isotopic mass of $^{243}$Am, 10.0 mg. However, this value does not agree with a spectroscopic characterization of the sample performed at CERN, which resulted into an $^{243}$Am mass of 7.34$\pm$15\% mg. An additional sample activity measurement performed with the TAC resulted into a mass of  6.77$\pm$15\% mg. The data were finally normalized to the transmission measurements available in EXFOR (see Section~\ref{sec:Normalization}), specially to the one performed by Simpson \textit{et al.} \cite{Simpson} , obtaining a normalization uncertainty of 3\%, and an associated sample mass of 6.23 ($\pm$4\%) mg, which is in agreement with the spectroscopic measurements. The impurities were determined during the resonance analysis process, finding around  0.048 mg of $^{241}$Am and 0.0025 mg of $^{240}$Pu. The temperature of the sample was assumed to be 293$\pm$4 K, which is the average temperature of the n\_TOF experimental area.

The sample was placed in the center of the TAC, held by two kapton foils of 25 $\mu$m thickness and surrounded by the neutron absorber. Due to the high sample activity, a Pb cylinder of 11.5 cm length and 1 mm thickness was placed around the sample, surrounding the 5.2 cm diameter vacuum tube. In this way, the amount of gamma rays with highest energies (200-300 keV) originated in the sample decay and reaching the TAC were strongly reduced. However, even with this lead shielding, the counting rate of this measurement was much higher than of other previous measurements performed with the TAC~\cite{Au197_Massimi,Np237_Guerrero}.

Three other samples were also measured for the determination of the $^{243}$Am(n,$\gamma$) cross section, with the same experimental conditions: (i) an empty Ti-Al canning similar to the one encapsulating the sample, with the same diameter but with a slightly different mass, 455.4(1)~mg, used for background determination purposes; (ii) a graphite sample of 10 mm diameter and 70.0(1) mg mass used to determine the TAC response to scattered neutrons; and (iii) a $^{197}$Au sample of 10 mm diameter and 185.4(1) mg mass used to determine the fraction of the beam intercepted by the $^{243}$Am sample, and also for validating the analysis tools.

Other measurements were also performed to determine the different background components: a measurement without beam and without sample in place (Env. Background), a measurement without beam and with the sample in place (Activity) and a measurement with neutron beam but without any sample (Empty frame). The time (pulses) and beam intensity (protons) allocated to each of these measurements is summarized in Table~\ref{tab:Meas_description}.

\begin{table}[bt]
\caption{\label{tab:Meas_description}
Number of pulses and protons dedicated to each measurement.
}
\begin{ruledtabular}
\begin{tabular}{lcc}
Measurement & \#pulses & \#protons \\
\colrule
\rule{0pt}{3ex}    
$^{243}$Am       & 1.86$\cdot$10$^{5}$ & 1.27$\cdot$10$^{18}$ \\
Env. background  & 1.37$\cdot$10$^{4}$ & -- \\
Activity         & 1.53$\cdot$10$^{4}$ & -- \\
$^{197}$Au       & 2.19$\cdot$10$^{4}$ & 1.53$\cdot$10$^{17}$ \\
Graphite         & 3.76$\cdot$10$^{3}$ & 2.67$\cdot$10$^{16}$ \\
Ti canning       & 1.49$\cdot$10$^{3}$ & 1.04$\cdot$10$^{16}$ \\
Empty frame      & 4.16$\cdot$10$^{3}$ & 2.94$\cdot$10$^{16}$ \\
\end{tabular}   
\end{ruledtabular}
\end{table}

\FloatBarrier
% =========================================================================================================

% =========================================================================================================
% =========================================================================================================
%                                             DATA REDUCTION
% =========================================================================================================

\section{Data reduction}{\label{sec:DataReduction}}

In this Section we describe the analysis process which leads to the experimental capture yield, which can be calculated as:
\begin{equation}
 Y_{n,\gamma}(E_{n})=\frac{C_{tot}(E_{n})-C_{bkg}(E_{n})}{\varepsilon \cdot F_{BIF} \cdot \phi(E_{n})}
\end{equation}

where $C_{tot}(E_{n})$ and $C_{bkg}(E_{n})$ are the number of total and background counts registered by the TAC, respectively, under certain  E$_{Sum}$ and m$_{cr}$ conditions; $\varepsilon$ is the detection efficiency under the same conditions; $\phi(E_{n})$ is the incident neutron fluence; and $F_{BIF}$ is the fraction of the neutron beam intercepted by the measured sample.

The data reduction process is quite similar to the one described in~\cite{Np237_Guerrero}, with some additional features specially developed to deal with the much higher counting rates (5.4 events/$\mu$s) observed in the $^{243}$Am(n,$\gamma$) measurement due to the sample activity.

\subsection{Background and selection of the analysis conditions}{\label{sec:Background}}

The background events in the $^{243}$Am(n,$\gamma$) measurement can be divided into two contributions: (i) events coming from fission reactions and scattered neutrons in the $^{243}$Am nuclei; and (ii) the rest of the background, which results from the environmental background, the activity of the BaF$_{2}$ crystals, the sample activity and the interaction of the neutron beam with all the materials except with the $^{243}$Am nuclei.

The second contribution could be obtained directly, in principle, from the different background measurements presented in Table~\ref{tab:Meas_description}, by subtracting and adding properly the different contributions. However, during the $^{243}$Am sample measurement the detection of the background events was distorted by the pile-up and dead time induced by the high $^{243}$Am sample activity, whereas in the background measurements it was not. This causes that the background can not be calculated directly from the dedicated background measurements and some corrections are needed. The procedure followed to take this effect into account is described in detail in~\cite{Mendoza_DT}, and it is based in the offline manipulation of the digitized detector signals and the parametrization of the response of the pulse shape analysis routine.

\begin{figure}[htb]
\begin{center}
\includegraphics[width=8.6cm]{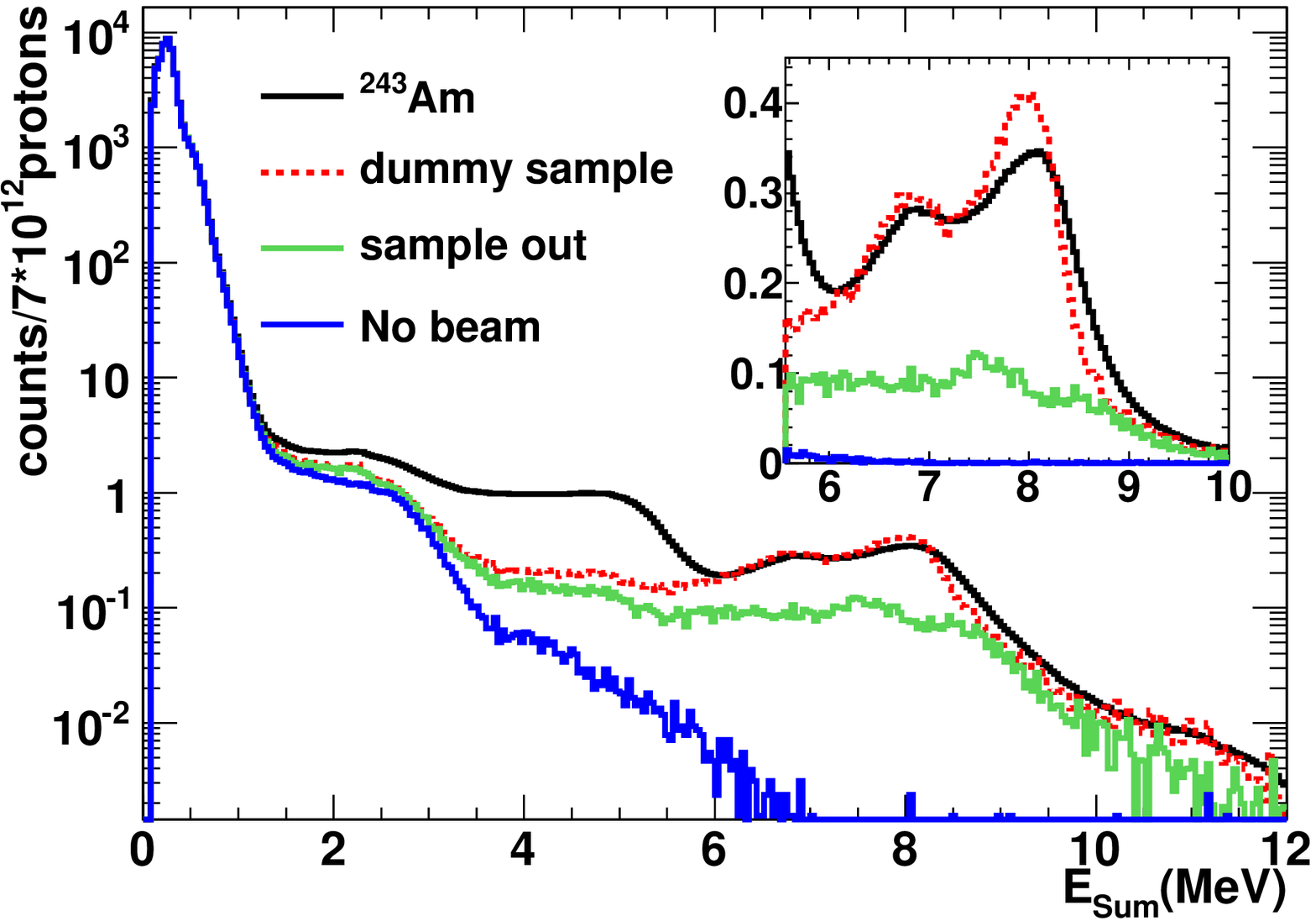} \\
\includegraphics[width=8.6cm]{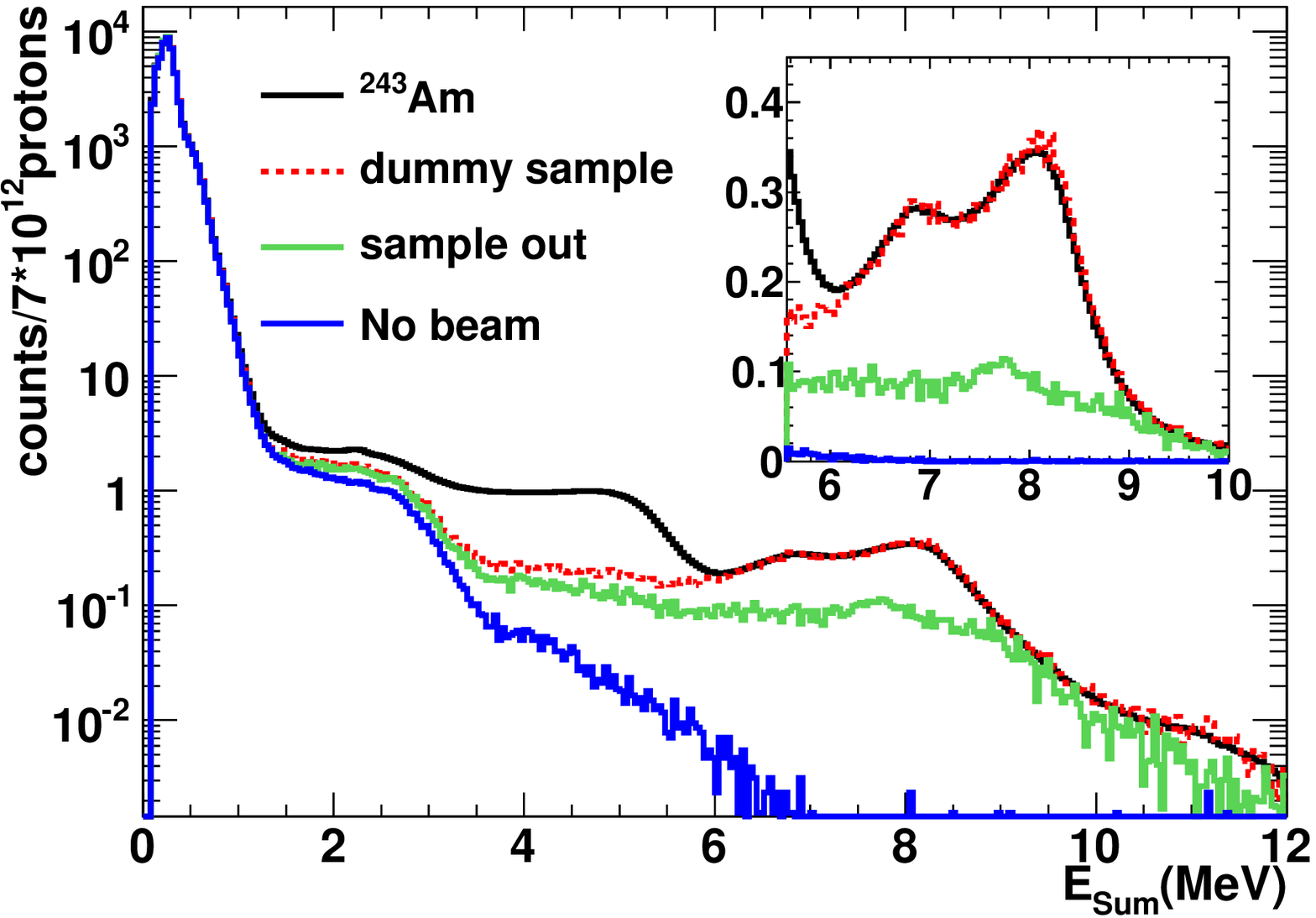}
\caption{(Color online) Deposited energy spectrum of the $^{243}$Am(n,$\gamma$) measurement together with different background contributions, without applying any condition in m$_{cr}$ and without (top) and with (bottom) applying pile-up corrections in the calculation of the backgrounds. A zoom of the high energy part is presented in the top-right corner of each panel. The data corresponds to neutron energies between 1 and 10 eV.}
\label{fig:Edep_Example}
\end{center}
\end{figure}

The deposited energy spectrum obtained from the $^{243}$Am(n,$\gamma$) measurement in the 1-10 eV neutron energy range is presented in Fig.~\ref{fig:Edep_Example}, together with different background contributions: the total contribution (dummy sample), the total contribution except the one related with the interaction of the neutron beam with the Ti capsule (sample out), and the contribution not related with the neutron beam (No beam). The part of the spectra below $\sim$2 MeV corresponds mostly to sample activity events, whereas above 6 MeV the events are due to background events related with the neutron beam, since the total energy of the $\gamma$-cascade emitted after a capture reaction in $^{243}$Am can not exceed the neutron separation energy of the compound nucleus, S$_{n}$($^{244}$Am)=5.36 MeV, and the no-beam background events have lower energies. For this reason, above E$_{Sum}$=6 MeV the dummy sample spectrum should match the results of the $^{243}$Am(n,$\gamma$) measurement. As it can be observed, this only happens if the mentioned pile-up corrections are applied (bottom panel).

\begin{figure}[htb]
\begin{center}
\includegraphics[width=8.6cm]{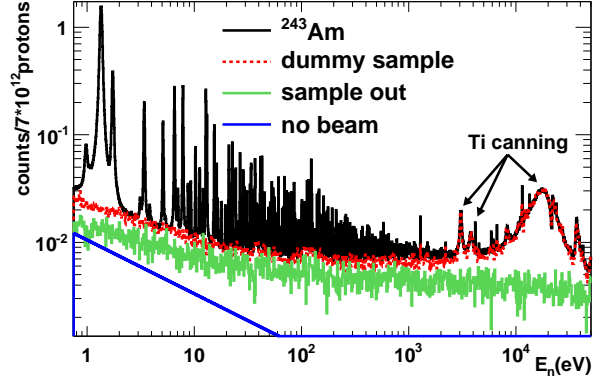}
\caption{(Color online) Number of events detected in the $^{243}$Am(n,$\gamma$) measurement as a function of the neutron energy, together with different background contributions and under the conditions of m$_{cr}>$2 and 2.5$<$E$_{Sum}<$6 MeV.}
\label{fig:CR_Example}
\end{center}
\end{figure}

It can also be observed that the capture to background ratio is highly improved if the low (E$_{Sum}<$2 MeV) and high (E$_{Sum}>$6 MeV) energy events are not considered. The same occurs if some conditions are applied on the m$_{cr}$, since the capture events have, in general, higher multiplicity than the background ones. On the other hand, the detection efficiency becomes lower as the conditions in E$_{Sum}$ and m$_{cr}$ are more restrictive. A detailed analysis has lead to the optimum conditions of m$_{cr}>$2 and 2.5$<$E$_{Sum}<$6 MeV adopted in the analysis. The number of events detected per proton pulse under these conditions is presented in Fig.~\ref{fig:CR_Example} as a function of the neutron energy. It can be appreciated that the background is smooth until E$_{n}$=2-3 keV, where the resonances of the Ti capsule appear. These Ti resonances have not allowed to measure above 2.5 keV, which is the high energy limit of this measurement. The low energy limit of 0.7 eV is given by the 16 ms recording time.

Due to small differences in the energy calibration caused by the pile-up correction method~\footnote{Data buffers of different measurements, which can have not exactly the same energy calibration, are added artificially to perform the pile-up corrections (see~\cite{Mendoza_DT})}, there was a background component constant in time that could not be determined from the measurements and had to be fitted. The uncertainty due to this fit can be expressed by considering the background $B(E_{n})$ as $B(E_{n})+a/\sqrt{E_{n}}$, where $a=0\pm3\cdot10^{-5}\sqrt{eV}$. The relative uncertainty of the background due to this component is 1\%, 0.6\%, 0.3\% and 0.13\% at 1, 10, 100 and 1000 eV, respectively.

The background contribution related with the interaction of neutrons in the $^{243}$Am nuclei follows the similar resonant behavior than the $^{243}$Am(n,$\gamma$) cross section. An estimation of this contribution can be performed with the evaluated cross sections if the probability of detecting a scattered neutron (neutron sensitivity, in this paper) and a fission reaction are known. The neutron sensitivity has been obtained from the measurement performed with the graphite sample (Table~\ref{tab:Meas_description}), by assuming that the neutrons scattered in Carbon have similar energies and angles than the neutrons scattered in $^{243}$Am. The neutron sensitivity depends on the neutron energy and also in the E$_{Sum}$ and m$_{cr}$ conditions considered. With the  conditions used in this analysis, the 2.2 MeV $\gamma$-rays resulting from neutron capture in the H of the neutron absorber are avoided, thus reducing the neutron sensitivity significantly. This calculated neutron sensitivity was used, together with the $^{243}$Am evaluated cross section present in the ENDF/B-VII.0 library, to estimate the background induced by the neutrons scattered in the AmO$_{2}$ sample, finding that its contribution to the total background is below 0.5\% in the entire energy range of the measurement, even in the peak of the resonances. For the fission events, if an overestimated detection efficiency of 100\% was assumed then its contribution to the total background would be higher than the previous one in the peak of certain $^{243}$Am resonances, but this fission contribution would always be below 1\% with respect to the capture yield. As a consequence, both contributions, elastic scattering and fission in the sample, have been neglected in the analysis.

\FloatBarrier

\subsection{Detection efficiency and determination of the sample activity}{\label{sec:Efficiency}}

The detection efficiency has been calculated from Monte Carlo simulations. The entire process starts with the generation of the electromagnetic cascades which follows the neutron capture, which has been performed with the DECAYGEN code~\cite{Tain_DecayGen}. The resulting cascades are then transported into the TAC geometry with a code based in the GEANT4 package~\cite{GEANT4}. In the last step the Monte Carlo results are reconstructed in the same way as it is done in a real experiment, including all the experimental effects such as the energy resolution of the crystals or the dead time and pile-up effects. The generation of the capture cascades includes statistical models for the description of the level densities and photon strength functions. These models depend on parameters, which are adjusted until the experimental results are reproduced. A detailed description of the entire process is given in~\cite{MC_TAC}, and this method has been also used in ~\cite{Np237_Guerrero}. The main difference introduced in this analysis is that a new dead time and pile-up correction method was developed, specially due to the strong effect of the high sample activity~\cite{Mendoza_DT}.

\begin{figure}[htb]
\begin{center}
\includegraphics[width=8.6cm]{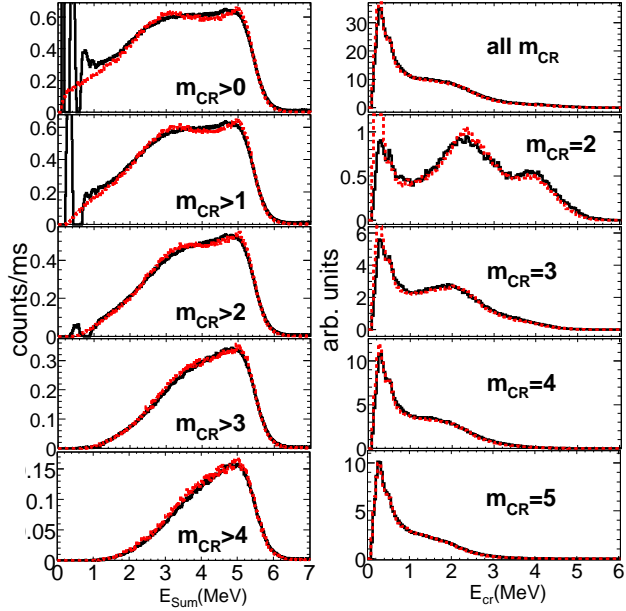}
\caption{(Color online) Experimental (solid lines) and simulated (dotted lines) deposited energy spectra from $^{243}$Am capture cascades, under different conditions in multiplicity. On the left, the total  $\gamma$-ray energy deposited in the TAC (E$_{Sum}$). On the right, the individual crystal energy spectra obtained by gating on the total $\gamma$-ray energy in the 4$<$E$_{Sum}<$6 MeV region. The results have been obtained from the strongest $^{243}$Am resonance at 1.35 eV.}
\label{fig:Efficiency}
\end{center}
\end{figure}

The mentioned statistical parameters have been adjusted to reproduce: (i) the deposited energy (E$_{Sum}$) spectra for different detection multiplicities (m$_{cr}$); and (ii) the individual $\gamma$-ray energy spectra contributing to events with 4$<$E$_{Sum}<$6 MeV, where most of the capture cascade has been detected. The experimental spectra have been obtained for the strongest $^{243}$Am resonance at 1.35 eV, where the capture to background ratio is maximum. An example of how the experimental results are well reproduced is shown in Fig.~\ref{fig:Efficiency}. 

We have not found any significant difference in the shape of the deposited energy spectra between several resonances, and thus it was assumed that the detection efficiency depends only on the analysis conditions in E$_{Sum}$ and m$_{cr}$ and in the detected counting rate, $CR$, due to the associated pile-up and dead time effects. Thus, $\varepsilon=\varepsilon({E_{Sum},m_{cr}},CR)$, and the variations in the detection efficiency with the neutron energy are only due to changes in the detected counting rate. The Monte Carlo simulations allow to determine the detection efficiency for any E$_{Sum}$ and m$_{cr}$ conditions, and for any detected counting rate. For the conditions used in this analysis, 2.5$<$E$_{Sum}<$6 MeV and m$_{cr}>$2, the calculated detection efficiency for low counting rates is 56.3(12)\%, and it varies less than 1\% due to the counting rate in the entire neutron energy range of the analysis. The estimation of the uncertainty in the efficiency was performed taking into account uncertainties in the generation of the cascades and uncertainties in the simulated TAC geometry. More details can be found in~\cite{MENDOZA_CGS,TesisEMendoza}.

\begin{figure}[htb]
\begin{center}
\includegraphics[width=8.6cm]{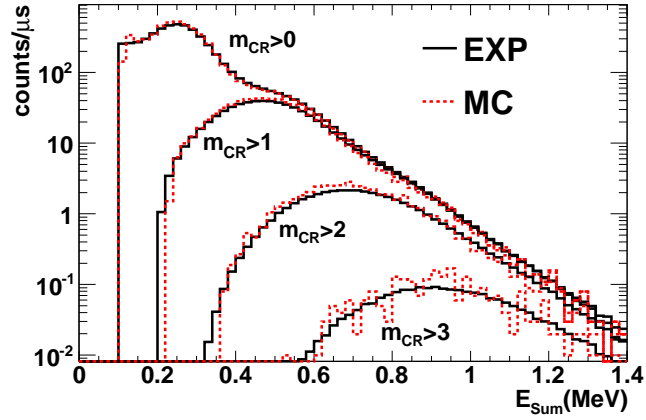}
\caption{(Color online) Experimental (solid lines) and simulated (dotted lines) deposited energy spectra due to the detection of the sample activity. A sample mass of 6.77 mg of $^{243}$Am has been used.}
\label{fig:Activity}
\end{center}
\end{figure}

The same tools used to calculate the detection efficiency were used to reproduce the energy response of the TAC to the sample activity. In this way, the value of the sample mass could be deduced by comparing the Monte Carlo simulations with the data. We obtained a sample mass of 6.77$\pm$15\% mg, which is consistent with the results of the spectroscopic characterization of the sample performed at CERN  (7.34$\pm$15\% mg) and not with the value provided by the manufacturers (10 mg). An example of the comparison between the experimental and the simulated results is given in Fig.~\ref{fig:Activity}. The estimated uncertainty is much larger than the one of the detection efficiency due to the lower energies of the $\gamma$-rays involved in the simulation, which in this case are close to the 100~keV threshold of the BaF$_{2}$ crystals.

\FloatBarrier

\subsection{Normalization}{\label{sec:Normalization}}

The fraction of the neutron beam intercepted by the measured sample, $F_{BIF}$, is the other quantity, together with the detection efficiency and the sample mass, which determines the normalization of the measurement. It has been calculated by measuring a thick $^{197}$Au sample of the same diameter as the  $^{243}$Am one (Section~\ref{sec:theSample}), placed at the same position. The strongest $^{197}$Au resonance at 4.9 eV has been used to measure the $F_{BIF}$ by means of the saturated resonance method~\cite{SRM_Macklin} obtaining a value of 0.196(3), which is consistent with other measured values~\cite{nTOFBeamProfile,Np237_Guerrero} for the same sample diameter.

The uncertainty in the normalization of the experimental capture yield is dominated by the uncertainties in the detection efficiency (2.2\%) and the $F_{BIF}$ (1.5\%), which added linearly or quadratically give total uncertainties of 3.7\% or 2.7\%, respectively. However, the uncertainty in the sample mass is much larger (11\%), so the n\_TOF capture measurement was finally normalized to the previous existing transmission measurements (Table~\ref{tab:differential_measurements}). The normalization procedure was performed with the SAMMY code~\cite{SAMMY}, by fitting the obtained capture yield to the existing transmission data. Two different methods were used:
\begin{enumerate}
 \item A simultaneous fit of the n\_TOF capture yield and the transmission measurements, where the resonance parameters and the normalization of the n\_TOF capture yield were varied. Only the Simpson \textit{et al.} data sets were used for these analyses, due to the lack of experimental information available for the rest of the transmission measurements, necessary to perform the fits.
 \item A normalization of the n\_TOF data to the resonance parameters provided by the experimentalists of the transmission measurements~\cite{Berreth,Simpson,Cote,Bellanova}, or the evaluators~\cite{Mughabghab_atlas2006,MASLOVEVALUATION}.
\end{enumerate}

The Simpson \textit{et al.} transmission measurement was performed with two $^{243}$Am samples, one thicker~\footnote{Simpson \textit{et al.}, file EXFOR 10204.004, retrieved from the IAEA Nuclear Data Services website.} than the other~\footnote{Simpson \textit{et al.}, file EXFOR 10204.005, retrieved from the IAEA Nuclear Data Services website.}. The normalization of the n\_TOF capture data was performed to both data sets in six different energy ranges~\footnote{The energy ranges considered were, in eV: 3-50, 3-25, 3-10, 8.5-12.5, 10-17 and 14-25.}. These transmission data were only used to normalize the capture data, and not to perform the resonance analysis, for two reasons. First, the uncertainties in the transmission data available in EXFOR are not given and thus realistic assumptions are necessary to perform the resonance analysis. It was estimated that reasonable assumptions can be made to perform a normalization, but are not sufficient to perform a resonance analysis. Second, the resolution function of the measurement was not reported and thus it had to be taken from a different reference. This is why the normalization was performed only at low neutron energies, below 50 eV, were the effect of the resolution function is very low.

In the second method we fitted the n\_TOF capture yield to the theoretical capture yield resulting from the different resonance parameters obtained by experimentalists and evaluators. We found that our data are incompatible with the values provided by Cote \textit{et al.} and Bellanova \textit{et al.} (see Table~\ref{tab:differential_measurements}), but are in a reasonable agreement with the resonance parameters provided by Simpson \textit{et al.}, Berreth \textit{et al.}, and some evaluations such as the ones performed by Mughabghab~\cite{Mughabghab_atlas2006} or Maslov~\cite{MASLOVEVALUATION}.

\begin{figure}[htb]
\begin{center}
\includegraphics[width=8.6cm]{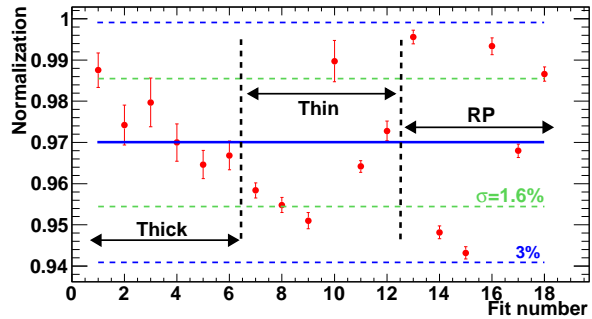}
\caption{(Color online) Results of 18 different normalization fits performed with SAMMY. In all the cases, the initial thickness of the $^{243}$Am sample considered was 2$\cdot$10$^{5}$ atoms/barn, which corresponds to a mass of 6.34 mg.}
\label{fig:Normalization}
\end{center}
\end{figure}

The results of all these normalization values are presented in Fig.~\ref{fig:Normalization}. The first 6 points correspond to the fits performed to the Simpson \textit{et al.} thick transmission sample, the fits 7 to 12 to the Simpson \textit{et al.} thin transmission sample, and the latter 6 points to the fits performed to the different resonance parameters. In all the cases the fits were performed above 3 eV, to avoid the strongest $^{243}$Am resonance at 1.35 eV, where the self-shielding corrections are relevant. The uncertainties in the normalization data points are due to the uncertainty in the background component constant in time presented in Section~\ref{sec:Background}, which is the dominant contribution. More information concerning the normalization procedure can be found in~\cite{TesisEMendoza}.

The mean value of all the normalization values presented in Fig.~\ref{fig:Normalization} is 0.970, which corresponds to a sample thickness of 1.94$\cdot$10$^{5}$ atoms/barn, or a mass of 6.23 mg of $^{243}$Am. The standard deviation is 1.6\%, but the different values are not independent and thus a 3\% uncertainty in the normalization was adopted, which is more conservative. Note that this 3\% uncertainty is the uncertainty in the normalization of the capture cross section. In the calculation of the uncertainty of the sample mass, the 2.7\% uncertainty in the normalization of the experimental capture yield (due to the detection efficiency and the $F_{BIF}$, without taking into account the normalization to transmission) has to be added. Thus, the fitted sample mass (or thickness) has an uncertainty of 4\%, if both quantities are added quadratically.

% =========================================================================================================

% =========================================================================================================
% =========================================================================================================
%                                             ANALYSIS
% =========================================================================================================
\FloatBarrier

\section{Cross section analysis}{\label{sec:Analysis}}

\FloatBarrier
\subsection{Analysis of the Resolved Resonance Region}{\label{sec:RRR}}

\begin{figure}[htb]
\begin{center}
\includegraphics[width=8.6cm]{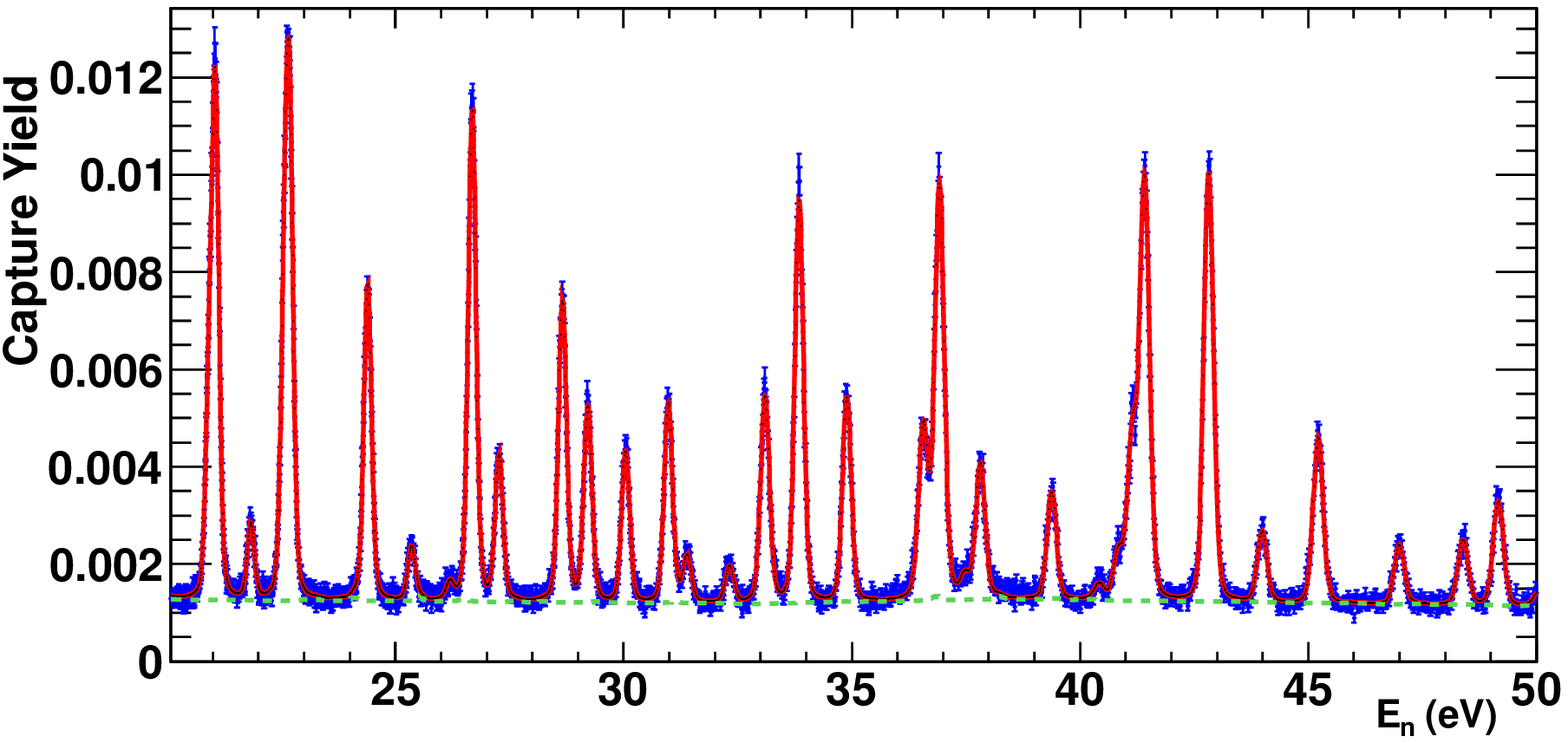} \\
\includegraphics[width=8.6cm]{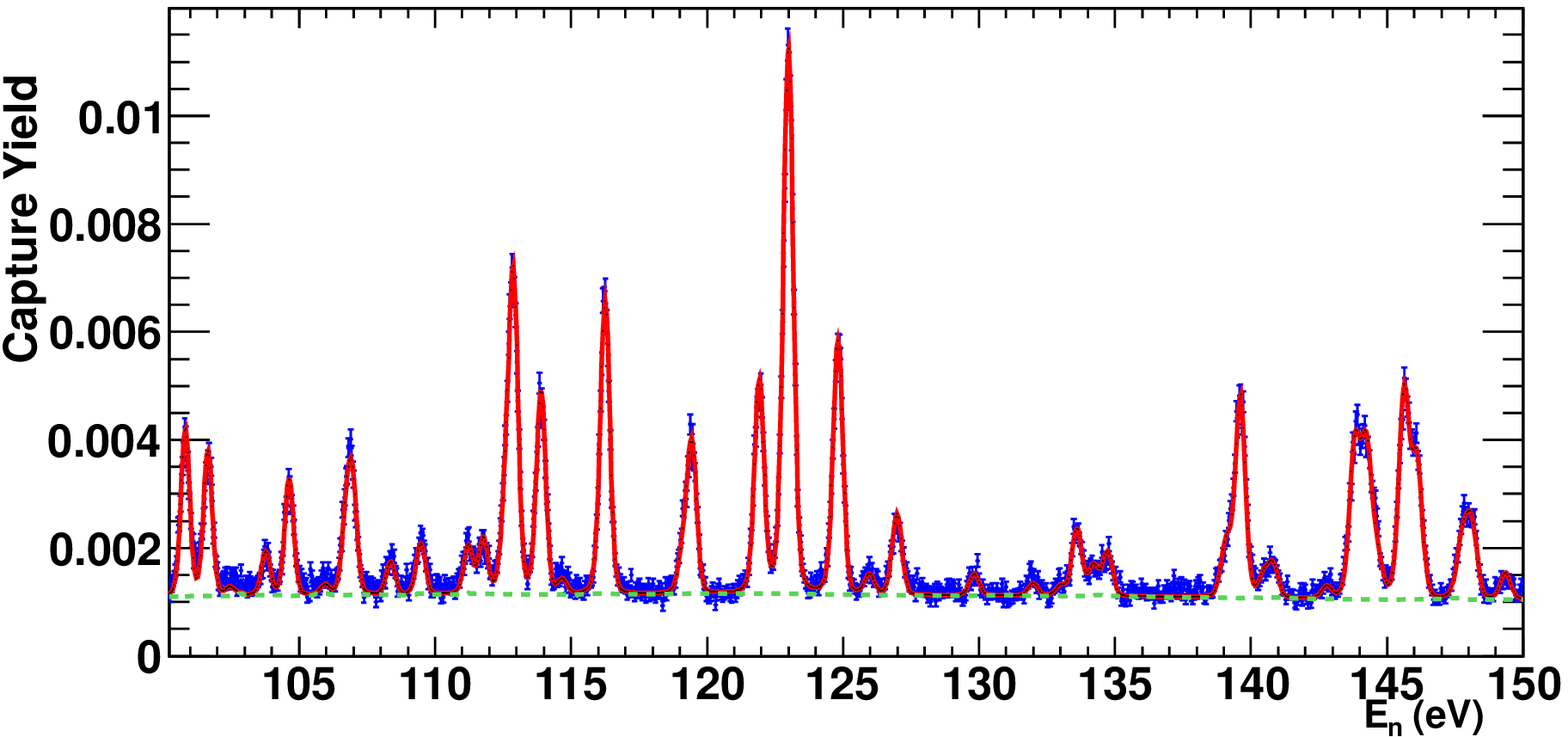} \\
\includegraphics[width=8.6cm]{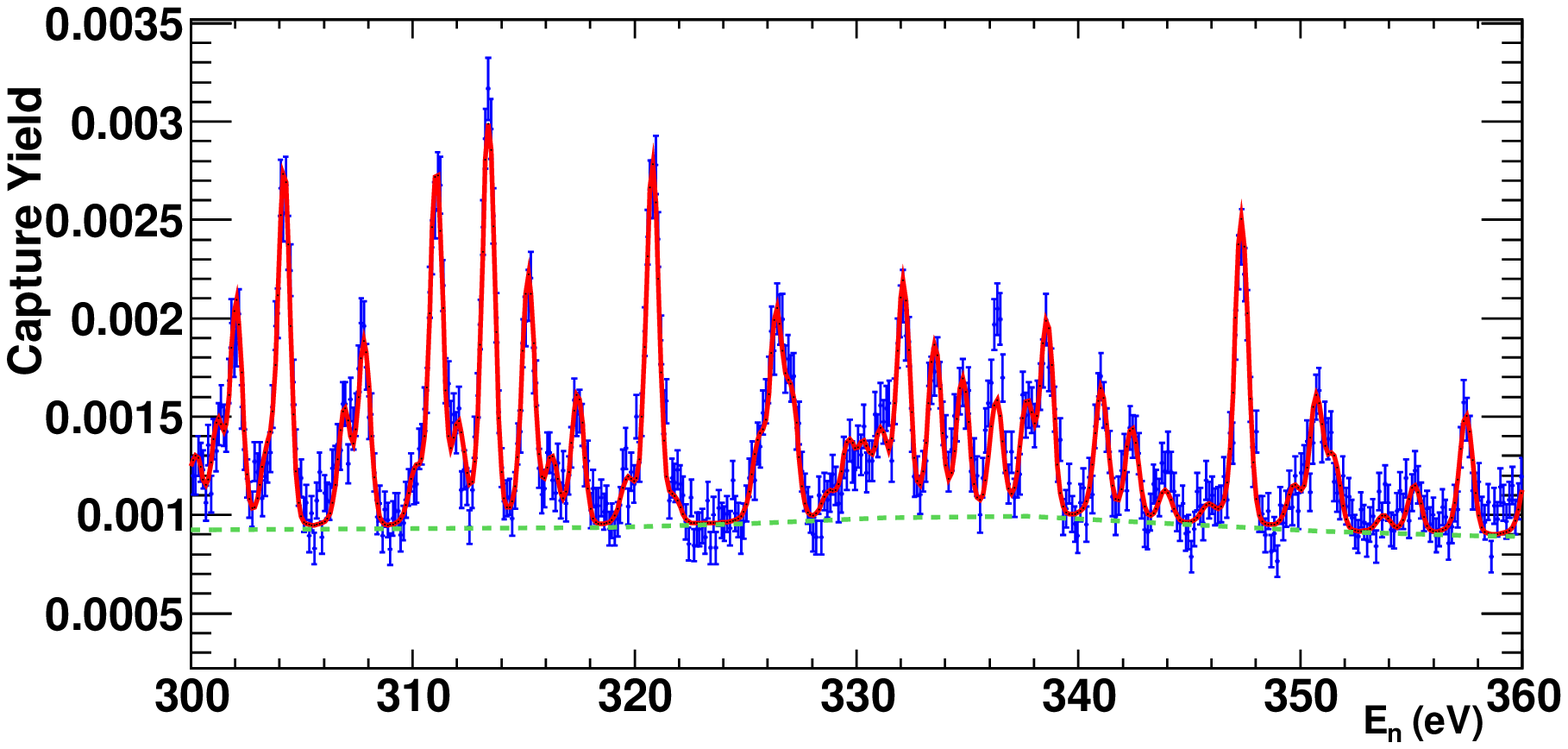}
\caption{(Color online) Examples of the fitted n\_TOF capture yield, in different energy ranges. The dashed line corresponds to the overall background considered in SAMMY.}
\label{fig:Yield_RRR}
\end{center}
\end{figure}

The Resolved Resonance Region (RRR) has been analyzed with the SAMMY code (version 7.0.0) up to 400 eV (250 eV is the high energy limit of the RRR in the present evaluations). We have fitted the energy $E_{0}$, the neutron width $\Gamma_{n}$, and the radiative capture width $\Gamma_{\gamma}$ of each resonance in the measured energy range, using the Reich-Moore approximation. The resonance parameters of the negative and the first resonance at 0.415 eV, the scattering radius, and all the fission widths were fixed to the values present in the ENDF/B-VII.0 evaluation, after verifying that strong variations of these parameters do not affect significantly the resulting capture yield. All the observed resonances are s-wave resonances (orbital spin $l=0$), as it was confirmed after performing the fit, by applying the techniques described in~\cite{SUGGEL}. It was not possible to distinguish between the two J=2,3 possible total spin values, so only the $g\Gamma_{n}$ values were determined. The time-energy relation was obtained by fitting the n\_TOF time of flight distance to reproduce the energies of the resonances of $^{197}$Au in the ENDF/B-VII.0 evaluation, obtaining 184.878 m. The n\_TOF capture yield is presented together with the results of the SAMMY fit in Fig.~\ref{fig:Yield_RRR}, for several neutron energy ranges.

We have obtained the statistical uncertainties in the resonance parameters from SAMMY, together with their correlations. Concerning the systematic uncertainties, the following contributions were considered:
\begin{enumerate}
 \item Uncertainties due to the normalization. They were estimated by performing several fits (1000), each of them with a different normalization value, varied randomly according to a Gaussian distribution with standard deviation equal to the 3\% uncertainty in the normalization. The systematic uncertainty of each fitted parameter was then estimated as the standard deviation of all the fitted values.
 \item Uncertainties due to the temperature of the sample. They were estimated in the same way as in the previous case, by varying the sample temperature according to 293$\pm$4 K.
 \item Uncertainties due to the background component constant in time (Section~\ref{sec:Background}). They were estimated in the same way as in the two previous cases, by varying this parameter according to its value: $a=0\pm3\cdot10^{-5}\sqrt{eV}$.
 \item Uncertainty due to the shape of the background. Due to the low beam time dedicated to the background measurements (Table~\ref{tab:Meas_description}), it was necessary to integrate the background in large neutron energy intervals to reduce the statistical fluctuations. However, since the shape of the background is quite smooth (Fig.~\ref{fig:CR_Example}), we used an smoothed background for the resonance analysis. In order to estimate the uncertainties in the resonance parameters due to the smoothing procedure, different analysis were performed, each of them with a background smoothed with a different technique. The uncertainties in the fitted resonance parameters were then estimated as the standard deviation of the resulting fitted values.
 \item Uncertainty due to the Doppler broadening model. Following the same approach than in the previous cases, we estimated this contribution by comparing the results of a fit performed with the free gas model and a fit performed with the crystal-lattice model~\cite{SAMMY}. In the latter case, we used the phonon spectrum of UO$_{2}$, since it has not been measured for AmO$_{2}$.
 \item Uncertainty due to the sample inhomogeneities. The resonance integral ($I_{0}=\int_{0.5eV}^{\infty}\sigma_{\gamma}(E)/EdE$) obtained after performing the fit is $I_{0}=$1681 barn, which is significantly lower than any of the measured values presented in Table~\ref{tab:integral_measurements}. This discrepancy can be explained with the existence of inhomogeneities in the sample, which would affect the shelf shielding and multiple scattering corrections. The strongest resonance at 1.35 eV is the only one where these corrections are important ($\sim$15\%), and this resonance contributes around 70-80\% to the resonance integral. In the rest of the resonances the shelf shielding and multiple scattering corrections are much lower (8 resonances with corrections between 3\% and 1\%, the rest of the resonances below 1\%). For this reason, the strongest resonance at 1.35 eV was not measured correctly. This is why the normalization to the transmission data was performed above 3 eV. In order to estimate the uncertainties due to the sample inhomogeneities, we compared the results of a fit performed with the nominal sample thickness with a fit performed with a double thickness, where the shelf shielding and multiple scattering corrections are larger.
\end{enumerate}

The rest of the contributions, such are the ones corresponding to the dead time corrections or the resolution function, were considered negligible.

The values of the fitted resonance parameters are presented in Tables \ref{tab:RP_07_50eV} and \ref{tab:RP_50_400eV}. The $\Gamma_{\gamma}$ values with statistical uncertainties larger than 10\% were fixed to the average radiative capture width, which was calculated from the rest of the values, all of them from resonances below 43 eV. Table \ref{tab:RP_07_50eV} provides as well the (quadratic) sum of the systematic uncertainties. In the case of the $g\Gamma_{n}$ parameters, the contributions to the systematic uncertainties associated with the temperature, the shape of the background and the Doppler broadening are negligible. In addition, since for nuclei with $\Gamma_{\gamma}\gg\Gamma_{n}$ the resonance area is nearly proportional to $g\Gamma_{n}$, the uncertainty in the $g\Gamma_{n}$ due to the normalization is the same 3\% as the normalization uncertainty and has not been included in the tables. Thus, only the uncertainties due to the background component constant in time and the sample inhomogeneities were taken into account in the tabulated values. Concerning the $\Gamma_{\gamma}$ parameters, the normalization is the only negligible contribution to the total systematic uncertainty.

Above 43 eV, all the $\Gamma_{\gamma}$ values were fixed to $<\Gamma_{\gamma}>=42$ meV, and only the energy and $g\Gamma_{n}$ values are given in Table \ref{tab:RP_50_400eV}. At these energies, the estimated uncertainties due to the background component constant in time and the sample inhomogeneities are negligible, so only the statistical uncertainty and the uncertainty due to the normalization have to be considered.

More information concerning the correlations between the different resonance parameters and the different contributions to the systematic uncertainties will be made available in EXFOR.

\LTcapwidth=8.6cm

\begin{longtable}{lcccccc}
\caption{\label{tab:RP_07_50eV}Resonance parameters below 43 eV, together with their statistical uncertainties ($\sigma_{stat}$), their total systematic uncertainties ($\sigma_{sys}$) in the case of the $\Gamma_{\gamma}$ parameters, and the sum of the systematic uncertainties with the exception of the one due to the normalization ($\sigma_{sys*}$) in the case of the $g\Gamma_{n}$ parameters. All the $g\Gamma_{n}$ values have an additional 3\% systematic uncertainty due to the normalization which has not been included in $\sigma_{sys*}$.} \\
\hline\hline
E$_{0}$ & $g\Gamma_{n}$ & $\sigma_{stat}$ & $\sigma_{sys*}$ & $\Gamma_{\gamma}$ & $\sigma_{stat}$ & $\sigma_{sys}$\\
 (eV) &  (meV) & (meV) & (meV) & (meV) & (meV)  & (meV) \\
\hline \endfirsthead
\caption{(continued)}\\ 
\hline\hline  
E$_{0}$ & $g\Gamma_{n}$ & $\sigma_{stat}$ & $\sigma_{sys*}$ & $\Gamma_{\gamma}$ & $\sigma_{stat}$ & $\sigma_{sys}$\\
 (eV) &  (meV) & (meV) & (meV) & (meV) & (meV)  & (meV) \\
\hline
\endhead
\hline\hline \\  
\endfoot
 -2 & 0.5735 & & & 39 \\
 0.4151 & 0.00042 & & & 39 \\
 0.9798 & 0.00643 & 0.00004 & 0.00008 & 34.4 & 0.4 & 1.1 \\
 1.3526 & 0.48579 & 0.00024 & 0.02447 & 48.57 & 0.04 & 2.50 \\
 1.7395 & 0.11465 & 0.00015 & 0.00111 & 40.11 & 0.12 & 0.39 \\
 3.1251 & 0.00486 & 0.00012 & 0.00015 & 34.0 & 2.4 & 1.8 \\
 3.4160 & 0.1389 & 0.0003 & 0.0007 & 39.93 & 0.24 & 0.47 \\
 3.8382 & 0.00608 & 0.00017 & 0.00022 & 45.2 & 2.9 & 2.4 \\
 5.1122 & 0.1512 & 0.0005 & 0.0007 & 40.2 & 0.4 & 0.6 \\
 6.5378 & 0.4824 & 0.0011 & 0.0031 & 41.0 & 0.3 & 0.7 \\
 7.0467 & 0.0359 & 0.0005 & 0.0004 & 47.8 & 2.1 & 1.3 \\
 7.8434 & 0.6813 & 0.0015 & 0.0048 & 42.9 & 0.3 & 0.8 \\
 8.3658 & 0.00788 & 0.00036 & 0.00016 & 42 \\
 8.7480 & 0.0630 & 0.0008 & 0.0005 & 45.5 & 1.8 & 1.2 \\
 9.2931 & 0.0745 & 0.0008 & 0.0006 & 40.3 & 1.6 & 1.3 \\
 10.286 & 0.2384 & 0.0013 & 0.0009 & 53.5 & 0.9 & 1.0 \\
 10.870 & 0.00769 & 0.00046 & 0.00016 & 42 \\
 11.249 & 0.1458 & 0.0013 & 0.0007 & 41.1 & 1.4 & 1.0 \\
 11.661 & 0.0518 & 0.0010 & 0.0004 & 40.8 & 2.8 & 0.8 \\
 12.098 & 0.0855 & 0.0012 & 0.0006 & 42.3 & 2.2 & 1.0 \\
 12.846 & 1.189 & 0.004 & 0.008 & 43.5 & 0.5 & 0.9 \\
 13.124 & 0.713 & 0.003 & 0.003 & 46.1 & 0.8 & 1.7 \\
 15.098 & 0.03001 & 0.00115 & 0.00019 & 42 \\
 15.369 & 0.6881 & 0.0032 & 0.0021 & 44.0 & 0.9 & 1.3 \\
 16.178 & 0.2725 & 0.0024 & 0.0010 & 44.9 & 1.6 & 1.2 \\
 16.543 & 0.0978 & 0.0015 & 0.0004 & 42 \\
 17.830 & 0.1115 & 0.0017 & 0.0005 & 42 \\
 18.120 & 0.0200 & 0.0011 & 0.0003 & 42 \\
 19.496 & 0.1089 & 0.0019 & 0.0006 & 42 \\
 19.869 & 0.0425 & 0.0015 & 0.0005 & 42 \\
 20.933 & 0.2356 & 0.0040 & 0.0006 & 42 \\
 21.071 & 0.5822 & 0.0054 & 0.0018 & 42.8 & 1.7 & 1.3 \\
 21.840 & 0.0657 & 0.0028 & 0.0004 & 42 \\
 21.851 & 0.02369 & 0.00224 & 0.00006 & 42 \\
 22.580 & 0.3509 & 0.0064 & 0.0011 & 41.8 & 2.7 & 0.8 \\
 22.695 & 0.5875 & 0.0070 & 0.0014 & 42.5 & 2.0 & 1.5 \\
 24.404 & 0.4761 & 0.0045 & 0.0017 & 50.6 & 2.0 & 1.4 \\
 25.365 & 0.0797 & 0.0024 & 0.0006 & 42 \\
 26.208 & 0.0257 & 0.0017 & 0.0003 & 42 \\
 26.695 & 0.8469 & 0.0062 & 0.0020 & 44.6 & 1.6 & 1.4 \\
 27.284 & 0.2452 & 0.0041 & 0.0011 & 46.2 & 3.3 & 0.9 \\
 28.673 & 0.5589 & 0.0057 & 0.0015 & 45.3 & 2.3 & 1.3 \\
 29.230 & 0.3650 & 0.0051 & 0.0012 & 45.3 & 2.9 & 1.2 \\
 30.058 & 0.2916 & 0.0043 & 0.0009 & 42 \\
 30.994 & 0.4023 & 0.0055 & 0.0014 & 42.6 & 2.9 & 1.4 \\
 31.406 & 0.0937 & 0.0033 & 0.0006 & 42 \\
 32.339 & 0.0768 & 0.0031 & 0.0007 & 42 \\
 33.115 & 0.4816 & 0.0064 & 0.0016 & 51.5 & 3.1 & 1.2 \\
 33.862 & 0.9399 & 0.0082 & 0.0020 & 40.6 & 2.0 & 1.8 \\
 34.908 & 0.4928 & 0.0064 & 0.0012 & 42 \\
 36.583 & 0.4657 & 0.0086 & 0.0014 & 64 & 4 & 3 \\
 36.940 & 1.1680 & 0.0113 & 0.0020 & 57 & 3 & 3 \\
 37.498 & 0.0532 & 0.0035 & 0.0003 & 42 \\
 37.837 & 0.3786 & 0.0071 & 0.0015 & 54.8 & 4.0 & 1.2 \\
 39.399 & 0.3007 & 0.0061 & 0.0011 & 42 \\
 40.438 & 0.0456 & 0.0033 & 0.0005 & 42 \\
 40.861 & 0.1369 & 0.0054 & 0.0007 & 42 \\
 41.165 & 0.4730 & 0.0096 & 0.0007 & 42 \\
 41.438 & 1.342 & 0.013 & 0.003 & 44.5 & 2.7 & 2.1 \\
 42.845 & 1.493 & 0.014 & 0.003 & 43.5 & 2.2 & 2.3 \\
%\hline \hline
%\end{tabular}   
%\end{ruledtabular}
\end{longtable}
%\end{center}

\begin{longtable}{lclc}
\caption{\label{tab:RP_50_400eV} Resonance energies and $g\Gamma_{n}$ parameters between 43 and 400 eV, the latter together with their statistical uncertainties ($\sigma_{stat}$). The systematic uncertainty in all the $g\Gamma_{n}$ values is the same, and amounts to a 3\% uncertainty due to the normalization.} \\
\hline\hline
E$_{0}$  & $g\Gamma_{n}\pm\sigma_{stat}$  & E$_{0}$ & $g\Gamma_{n}\pm\sigma_{stat}$ \\
 (eV) & (meV)  & (eV) & (meV)  \\

\hline \endfirsthead
\caption{(continued)}\\ 
\hline\hline
E$_{0}$ (eV) & $g\Gamma_{n}\pm\sigma_{stat}$(meV)  & E$_{0}$ (eV) & $g\Gamma_{n}\pm\sigma_{stat}$(meV)  \\
\hline
\endhead
\hline\hline \\  
\endfoot
 44.016 & 0.2311 $\pm$ 0.0064 & 45.242 & 0.5887 $\pm$ 0.0092\\
 47.018 & 0.2125 $\pm$ 0.0067 & 48.418 & 0.2416 $\pm$ 0.0073\\
 49.189 & 0.4093 $\pm$ 0.0088 & 50.107 & 0.058 $\pm$ 0.006 \\
 51.144 & 0.507 $\pm$ 0.010  &  52.026 & 0.039 $\pm$ 0.006  \\
 52.916 & 1.043 $\pm$ 0.014  &  53.579 & 0.035 $\pm$ 0.007  \\
 53.868 & 0.325 $\pm$ 0.010  &  54.393 & 0.871 $\pm$ 0.015  \\
 54.617 & 0.153 $\pm$ 0.011  &  55.737 & 0.912 $\pm$ 0.014  \\
 57.194 & 0.051 $\pm$ 0.007  &  58.572 & 0.200 $\pm$ 0.010  \\
 58.953 & 0.447 $\pm$ 0.012  &  59.803 & 0.444 $\pm$ 0.012  \\
 60.605 & 0.590 $\pm$ 0.014  &  61.049 & 1.527 $\pm$ 0.020  \\
 62.370 & 0.139 $\pm$ 0.009  &  63.032 & 0.226 $\pm$ 0.010  \\
 63.489 & 0.067 $\pm$ 0.009  &  64.664 & 0.314 $\pm$ 0.012  \\
 66.067 & 0.728 $\pm$ 0.015  &  67.194 & 0.602 $\pm$ 0.015  \\
 67.836 & 0.644 $\pm$ 0.015  &  68.524 & 0.870 $\pm$ 0.018  \\
 69.502 & 1.93 $\pm$ 0.03  &  70.102 & 1.402 $\pm$ 0.023  \\
 71.530 & 0.094 $\pm$ 0.011  &  72.035 & 1.392 $\pm$ 0.023  \\
 72.711 & 1.647 $\pm$ 0.024  &  73.713 & 0.235 $\pm$ 0.013  \\
 74.131 & 0.266 $\pm$ 0.014  &  74.785 & 0.111 $\pm$ 0.012  \\
 75.253 & 1.67 $\pm$ 0.03  &  76.385 & 0.158 $\pm$ 0.013  \\
 76.818 & 0.301 $\pm$ 0.015  &  77.362 & 0.832 $\pm$ 0.020  \\
 78.040 & 0.212 $\pm$ 0.014  &  79.821 & 0.114 $\pm$ 0.013  \\
 80.396 & 0.290 $\pm$ 0.022  &  80.611 & 0.54 $\pm$ 0.03  \\
 80.899 & 1.75 $\pm$ 0.03  &  82.862 & 0.410 $\pm$ 0.018  \\
 83.309 & 1.37 $\pm$ 0.03  &  83.963 & 0.27 $\pm$ 0.07  \\
 84.011 & 0.79 $\pm$ 0.08  &  84.599 & 0.227 $\pm$ 0.016  \\
 85.267 & 0.92 $\pm$ 0.06  &  85.391 & 2.59 $\pm$ 0.07  \\
 86.436 & 0.983 $\pm$ 0.024  &  88.133 & 0.725 $\pm$ 0.023  \\
 88.740 & 0.819 $\pm$ 0.024  &  90.165 & 0.697 $\pm$ 0.023  \\
 91.002 & 0.607 $\pm$ 0.022  &  94.474 & 0.75 $\pm$ 0.03  \\
 95.081 & 0.105 $\pm$ 0.017  &  95.642 & 0.252 $\pm$ 0.020  \\
 97.283 & 1.10 $\pm$ 0.03  &  98.509 & 0.218 $\pm$ 0.020  \\
 99.280 & 0.402 $\pm$ 0.023  &  100.870 & 1.83 $\pm$ 0.04  \\
 101.701 & 1.53 $\pm$ 0.04  &  102.521 & 0.212 $\pm$ 0.021  \\
 103.830 & 0.44 $\pm$ 0.03  &  104.690 & 1.26 $\pm$ 0.03  \\
 106.032 & 0.182 $\pm$ 0.022  &  106.793 & 0.96 $\pm$ 0.05  \\
 107.022 & 1.12 $\pm$ 0.05  &  108.425 & 0.43 $\pm$ 0.03  \\
 109.531 & 0.68 $\pm$ 0.03  &  111.280 & 0.61 $\pm$ 0.03  \\
 111.831 & 0.60 $\pm$ 0.03  &  112.651 & 1.14 $\pm$ 0.05  \\
 112.945 & 4.20 $\pm$ 0.08  &  113.955 & 2.79 $\pm$ 0.06  \\
 114.731 & 0.24 $\pm$ 0.03  &  116.120 & 0.14 $\pm$ 0.09  \\
 116.316 & 4.18 $\pm$ 0.04  &  119.186 & 0.62 $\pm$ 0.04  \\
 119.507 & 2.04 $\pm$ 0.06  &  121.983 & 3.14 $\pm$ 0.06  \\
 123.055 & 14.1 $\pm$ 0.3\hspace{5pt} &  124.880 & 4.07 $\pm$ 0.08  \\
 126.037 & 0.34 $\pm$ 0.03  &  127.053 & 1.13 $\pm$ 0.04  \\
 129.891 & 0.32 $\pm$ 0.03  &  132.060 & 0.31 $\pm$ 0.03  \\
 133.064 & 0.18 $\pm$ 0.03  &  133.669 & 1.05 $\pm$ 0.05  \\
 134.271 & 0.43 $\pm$ 0.04  &  134.795 & 0.65 $\pm$ 0.04  \\
 139.162 & 0.84 $\pm$ 0.05  &  139.682 & 4.07 $\pm$ 0.09  \\
 140.529 & 0.30 $\pm$ 0.04  &  140.871 & 0.53 $\pm$ 0.05  \\
 142.862 & 0.28 $\pm$ 0.04  &  143.893 & 2.80 $\pm$ 0.08  \\
 144.317 & 2.75 $\pm$ 0.10  &  144.701 & 0.95 $\pm$ 0.06  \\
 145.705 & 4.35 $\pm$ 0.11  &  146.182 & 2.78 $\pm$ 0.08  \\
 147.839 & 1.16 $\pm$ 0.07  &  148.190 & 1.44 $\pm$ 0.07  \\
 149.436 & 0.43 $\pm$ 0.04  &  150.717 & 0.44 $\pm$ 0.04  \\
 152.491 & 0.71 $\pm$ 0.05  &  153.616 & 2.19 $\pm$ 0.08  \\
 154.267 & 1.86 $\pm$ 0.07  &  155.065 & 0.52 $\pm$ 0.05  \\
 158.180 & 1.85 $\pm$ 0.07  &  158.815 & 0.50 $\pm$ 0.05  \\
 160.229 & 6.05 $\pm$ 0.15  &  160.612 & 0.96 $\pm$ 0.07  \\
 163.471 & 0.29 $\pm$ 0.04  &  164.396 & 2.66 $\pm$ 0.08  \\
 165.683 & 0.64 $\pm$ 0.06  &  166.120 & 1.01 $\pm$ 0.07  \\
 166.469 & 0.35 $\pm$ 0.06  &  167.567 & 3.80 $\pm$ 0.11  \\
 169.394 & 0.73 $\pm$ 0.05  &  172.200 & 3.60 $\pm$ 0.10  \\
 173.081 & 4.40 $\pm$ 0.13  &  174.257 & 2.12 $\pm$ 0.08  \\
 175.280 & 1.91 $\pm$ 0.08  &  176.326 & 1.89 $\pm$ 0.10  \\
 176.727 & 3.13 $\pm$ 0.11  &  179.537 & 1.39 $\pm$ 0.09  \\
 179.911 & 0.98 $\pm$ 0.08  &  180.470 & 0.63 $\pm$ 0.06  \\
 181.226 & 1.19 $\pm$ 0.07  &  182.516 & 0.75 $\pm$ 0.06  \\
 183.579 & 1.82 $\pm$ 0.09  &  184.070 & 2.21 $\pm$ 0.09  \\
 185.608 & 0.65 $\pm$ 0.06  &  186.227 & 1.49 $\pm$ 0.09  \\
 186.654 & 1.22 $\pm$ 0.08  &  187.517 & 4.42 $\pm$ 0.14  \\
 188.382 & 0.62 $\pm$ 0.06  &  189.884 & 0.69 $\pm$ 0.08  \\
 190.250 & 0.67 $\pm$ 0.08  &  191.064 & 2.04 $\pm$ 0.10  \\
 191.783 & 2.45 $\pm$ 0.10  &  192.902 & 5.38 $\pm$ 0.17  \\
 195.077 & 0.24 $\pm$ 0.05  &  195.821 & 0.72 $\pm$ 0.07  \\
 196.473 & 1.11 $\pm$ 0.08  &  197.187 & 3.01 $\pm$ 0.12  \\
 199.272 & 2.53 $\pm$ 0.11  &  201.999 & 0.52 $\pm$ 0.07  \\
 202.586 & 0.20 $\pm$ 0.05  &  204.043 & 0.82 $\pm$ 0.08  \\
 204.673 & 1.17 $\pm$ 0.08  &  206.078 & 0.92 $\pm$ 0.08  \\
 207.572 & 1.63 $\pm$ 0.09  &  208.859 & 2.14 $\pm$ 0.11  \\
 210.283 & 1.67 $\pm$ 0.13  &  210.640 & 2.05 $\pm$ 0.17  \\
 211.071 & 3.47 $\pm$ 0.17  &  212.793 & 0.36 $\pm$ 0.07  \\
 213.932 & 3.47 $\pm$ 0.14  &  216.325 & 1.25 $\pm$ 0.09  \\
 219.483 & 1.39 $\pm$ 0.10  &  220.063 & 0.78 $\pm$ 0.10  \\
 220.603 & 0.99 $\pm$ 0.10  &  222.118 & 0.41 $\pm$ 0.08  \\
 221.197 & 0.68 $\pm$ 0.08  &  222.656 & 0.42 $\pm$ 0.08\\
 224.599 & 4.32 $\pm$ 0.17  &  225.471 & 1.82 $\pm$ 0.11  \\
 226.315 & 0.50 $\pm$ 0.09  &  226.798 & 1.49 $\pm$ 0.11  \\
 228.388 & 0.44 $\pm$ 0.07  &  232.332 & 4.45 $\pm$ 0.17  \\
 233.502 & 5.88 $\pm$ 0.23  &  235.447 & 0.65 $\pm$ 0.09  \\
 236.908 & 1.87 $\pm$ 0.12  &  238.468 & 1.08 $\pm$ 0.10  \\
 239.061 & 0.75 $\pm$ 0.12  &  239.468 & 1.06 $\pm$ 0.12  \\
 241.136 & 0.59 $\pm$ 0.09  &  242.239 & 2.10 $\pm$ 0.13  \\
 243.670 & 0.84 $\pm$ 0.25  &  243.781 & 1.3 $\pm$ 0.3  \\
 244.558 & 1.20 $\pm$ 0.11  &  246.467 & 3.24 $\pm$ 0.16  \\
 247.913 & 5.54 $\pm$ 0.21  &  248.655 & 2.32 $\pm$ 0.15  \\
 251.053 & 2.88 $\pm$ 0.16  &  252.214 & 6.0 $\pm$ 0.3  \\
 254.482 & 0.90 $\pm$ 0.11  &  255.742 & 12.9 $\pm$ 0.6\hspace{5pt}  \\
 256.329 & 1.12 $\pm$ 0.13  &  257.622 & 1.53 $\pm$ 0.13  \\
 258.561 & 1.49 $\pm$ 0.13  &  259.384 & 7.9 $\pm$ 0.3  \\
 260.652 & 2.70 $\pm$ 0.16  &  262.868 & 0.58 $\pm$ 0.10  \\
 265.358 & 1.23 $\pm$ 0.14  &  265.951 & 3.28 $\pm$ 0.22  \\
 266.598 & 5.3 $\pm$ 0.3  &  267.837 & 1.64 $\pm$ 0.14  \\
 271.683 & 5.9 $\pm$ 0.3  &  272.757 & 1.02 $\pm$ 0.12  \\
 273.987 & 7.4 $\pm$ 0.3  &  275.076 & 2.55 $\pm$ 0.17  \\
 276.909 & 1.35 $\pm$ 0.14  &  277.563 & 1.57 $\pm$ 0.15  \\
 278.924 & 2.91 $\pm$ 0.19  &  280.013 & 2.55 $\pm$ 0.18  \\
 280.910 & 0.51 $\pm$ 0.11  &  281.542 & 1.10 $\pm$ 0.15  \\
 282.317 & 3.8 $\pm$ 0.3  &  282.897 & 7.8 $\pm$ 0.4  \\
 285.633 & 1.18 $\pm$ 0.13  &  288.117 & 4.23 $\pm$ 0.24  \\
 289.485 & 4.9 $\pm$ 0.3  &  291.076 & 6.2 $\pm$ 0.3  \\
 295.672 & 3.39 $\pm$ 0.22  &  298.124 & 1.76 $\pm$ 0.16  \\
 299.693 & 0.97 $\pm$ 0.14  &  300.418 & 0.88 $\pm$ 0.14  \\
 301.363 & 1.43 $\pm$ 0.16  &  302.211 & 3.65 $\pm$ 0.24  \\
 303.571 & 0.66 $\pm$ 0.14  &  304.369 & 8.1 $\pm$ 0.4  \\
 307.084 & 1.86 $\pm$ 0.18  &  307.989 & 3.52 $\pm$ 0.24  \\
 310.268 & 0.69 $\pm$ 0.14  &  311.234 & 8.3 $\pm$ 0.4  \\
 312.223 & 1.40 $\pm$ 0.17  &  313.586 & 11.0 $\pm$ 0.6\hspace{5pt}  \\
 315.387 & 4.7 $\pm$ 0.3  &  316.432 & 0.93 $\pm$ 0.15  \\
 317.607 & 2.25 $\pm$ 0.20  &  319.832 & 0.68 $\pm$ 0.14  \\
 320.986 & 9.6 $\pm$ 0.5  &  321.949 & 0.81 $\pm$ 0.15  \\
 325.731 & 0.74 $\pm$ 0.20 & 325.890 & 0.47 $\pm$ 0.22 \\
 326.558 & 4.0 $\pm$ 0.3  &  327.256 & 2.7 $\pm$ 0.3  \\
 328.915 & 0.40 $\pm$ 0.13  &  329.789 & 0.90 $\pm$ 0.18  \\
 330.480 & 0.96 $\pm$ 0.19  &  331.261 & 1.84 $\pm$ 0.22  \\
 332.280 & 4.7 $\pm$ 0.3  &  333.703 & 3.2 $\pm$ 0.3  \\
 334.963 & 2.8 $\pm$ 0.3  &  336.479 & 4.4 $\pm$ 0.3  \\
 337.849 & 1.92 $\pm$ 0.23  &  338.761 & 4.0 $\pm$ 0.3  \\
 341.211 & 2.55 $\pm$ 0.25  &  342.592 & 1.62 $\pm$ 0.21  \\
 344.105 & 1.16 $\pm$ 0.19  &  346.021 & 0.45 $\pm$ 0.14  \\
 347.545 & 7.5 $\pm$ 0.5  &  349.886 & 0.62 $\pm$ 0.16  \\
 350.889 & 2.4 $\pm$ 0.3  &  351.673 & 1.35 $\pm$ 0.20  \\
 353.935 & 0.47 $\pm$ 0.14  &  355.336 & 1.08 $\pm$ 0.18  \\
 357.665 & 2.44 $\pm$ 0.24  &  360.478 & 2.26 $\pm$ 0.24  \\
 361.742 & 4.4 $\pm$ 0.4  &  362.279 & 3.8 $\pm$ 0.4  \\
 363.254 & 1.47 $\pm$ 0.21  &  364.130 & 1.35 $\pm$ 0.22  \\
 364.859 & 1.50 $\pm$ 0.22  &  367.281 & 2.8 $\pm$ 0.3  \\
 368.092 & 1.74 $\pm$ 0.24  &  369.593 & 22.2 $\pm$ 1.6\hspace{5pt}  \\
 370.875 & 3.6 $\pm$ 0.3  &  372.679 & 2.5 $\pm$ 0.3  \\
 373.365 & 1.09 $\pm$ 0.21  &  375.489 & 1.42 $\pm$ 0.21  \\
 376.691 & 2.4 $\pm$ 0.3  &  378.523 & 1.05 $\pm$ 0.22  \\
 379.189 & 3.9 $\pm$ 0.4  &  380.270 & 4.9 $\pm$ 0.4  \\
 381.400 & 3.1 $\pm$ 0.3  &  382.246 & 2.4 $\pm$ 0.3  \\
 384.143 & 2.9 $\pm$ 0.3  &  384.967 & 1.84 $\pm$ 0.25  \\
 388.234 & 3.3 $\pm$ 0.3  &  389.368 & 3.1 $\pm$ 0.3  \\
 390.358 & 0.71 $\pm$ 0.18  &  391.073 & 0.98 $\pm$ 0.20  \\
 392.314 & 1.66 $\pm$ 0.24  &  393.751 & 8.1 $\pm$ 0.6  \\
 395.109 & 0.53 $\pm$ 0.16  &  396.471 & 3.3 $\pm$ 0.4  \\
 396.987 & 1.5 $\pm$ 0.3  &  399.229 & 3.4 $\pm$ 0.3  \\
\end{longtable}

\begin{figure}[htb]
\begin{center}
\includegraphics[width=8.6cm]{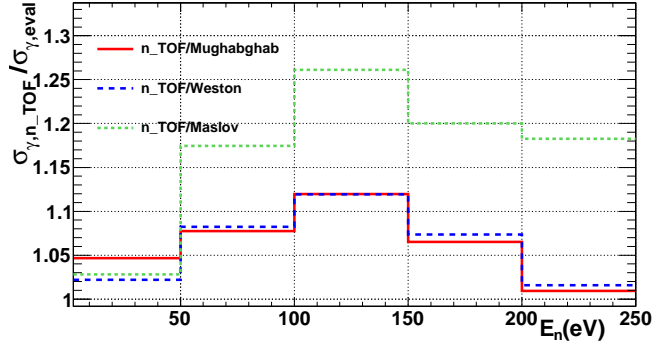} 
\caption{(Color online) Ratio between the n\_TOF fitted capture cross section and the ones available in different evaluations ($\int_{E_{1}}^{E_{2}}\sigma_{\gamma,n\_TOF}(E)dE/\int_{E_{1}}^{E_{2}}\sigma_{\gamma,eval}(E)dE$), integrated in different energy ranges. The low energy limit of the first bin is 3 eV, in order to avoid the strongest resonance at 1.35 eV.}
\label{fig:Comp_RRR}
\end{center}
\end{figure}

The ratio between the n\_TOF capture cross section and the most recent evaluations is presented in Fig.~\ref{fig:Comp_RRR}. The ratio has been performed with three different evaluations, carried out by: Mughabghab, adopted by the ENDF/B-VII.1 evaluation; Weston, adopted by the older ENDF/B releases and similar to the one by Mughabghab; and Maslov, adopted by the rest of the evaluations in this energy range: JEFF-3.1.2, JENDL-4.0, JENDL-3.3~\cite{JENDL-3.3}, BROND-2.2~\cite{BROND-2.2} and CENDL-3.1. In the 3-250 eV energy range, the n\_TOF capture cross section is, on average, 6\% larger than the Mughabghab and Weston evaluations and 13\% larger than the Maslov evaluation. It should be said that new resonances have been found in this energy range, as well as 105 new resonances between 250 and 400 eV. In particular, the present evaluations contain 218 (Mughabghab and Weston) and 238 (Maslov) resonances up to 250 eV, whereas there are 248 resonances in the n\_TOF results.

\FloatBarrier
\subsection{Statistical analysis of the resonance parameters}{\label{sec:StatParam}}

The average radiation width $\langle \Gamma_{\gamma} \rangle$ was determined from the fitted values available in Table~\ref{tab:RP_07_50eV}. We used the generalized weighted mean to take into account the correlations between the different parameters, but very similar results are obtained if the correlations are neglected. The resonances below 3 eV were not used to calculate $\langle \Gamma_{\gamma} \rangle$, due to the mentioned problems in the vicinity of the strongest resonance at 1.35 eV associated with the inhomogeneities. The resulting value was $\langle \Gamma_{\gamma} \rangle$=42.00$\pm$0.12$\pm$0.5$\pm$0.3$\pm$0.7$\pm$0.6 meV, where these uncertainties are, respectively: statistical, due to the sample temperature, due to the background component constant in time, due to the Doppler broadening model and due to the sample inhomogeneities. If all the systematic uncertainties are added linearly or quadratically we obtain a total systematic uncertainty of 2.1 or 1.1 meV, respectively.

\begin{figure}[htb]
\begin{center}
\includegraphics[width=8.6cm]{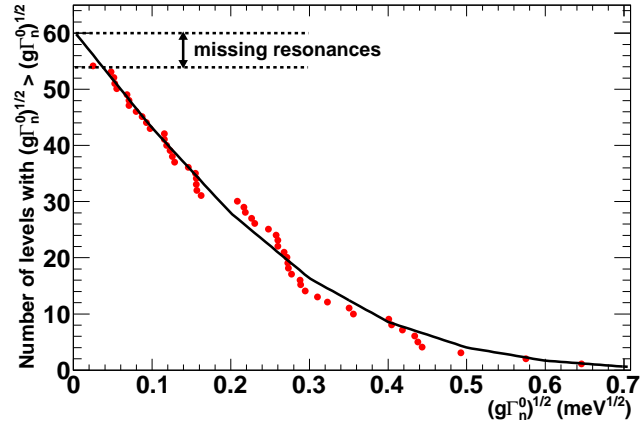} 
\caption{(Color online) Estimation of the number of missing resonances, performed in the 0-40 eV energy range. The experimental points were fitted to Equation~\ref{eq:MissingResonances}.}
\label{fig:MissingResonances}
\end{center}
\end{figure}

An estimation of the s-wave average level spacing $D_{0}$ can be obtained, in principle, from $D_{0}=\Delta E/(N-1)$ and $\Delta D_{0}/D_{0}\approx1/N$~\cite{RIPL3}, where $N$ is the number of resonances observed in the neutron energy interval between $E_{1}$ and $E_{2}$ and $\Delta E=E_{2}-E_{1}$. However, there are usually a certain number of small resonances which have not been detected (missing resonances), and their number has to be estimated as well. One of the most common used methods is based on assuming that the values of the reduced neutron widths $\Gamma_{n}^{0}=\Gamma_{n}\cdot (E_{0}/1eV)^{-1/2}$ are distributed, for each spin value $J$, according to a Porter-Thomas distribution with one degree of freedom $p(x)dx=e^{-x/2}/\sqrt{2\pi x}$, were $x=\Gamma_{n,J}^{0}/\langle \Gamma_{n,J}^{0} \rangle$. Since the spins of the resonances have not been determined, we assumed that $1/D_{0,J}\propto (2J+1)$ and that $S_{0,J}=S_{0}$ (both assumptions are justified in Appendix D of \cite{ENDF6MANUAL}), where $S_{0}=\langle g\Gamma_{n}^{0} \rangle /D_{0}$ is the s-wave neutron strength function. From these assumptions it follows that $\langle g_{J}\Gamma_{n,J}^{0} \rangle =\langle g\Gamma_{n}^{0} \rangle$, i.e., the value of $\langle g\Gamma_{n}^{0} \rangle$ is the same for both spin groups. Therefore, it is possible to consider only one Porter-Thomas distribution, where both spin groups are included, after making the change of variable from $x=\Gamma_{n}^{0}/\langle \Gamma_{n}^{0} \rangle$ to $y=g\Gamma_{n}^{0}/\langle g\Gamma_{n}^{0} \rangle$. With some manipulations of the Porter-Thomas distribution, it follows that, for a given energy interval, the number of resonances with $\sqrt{g\Gamma_{n}^{0}}$ greater than a certain value, $x$, is obtained from:
\begin{equation}
f(x)=N\frac{2}{\sqrt{\pi}\sqrt{2\langle g\Gamma_{n}^{0}\rangle}}\int_{x}^{\infty}{\exp \left(-\frac{y^{2}}{2\langle g\Gamma_{n}^{0}\rangle}\right) dy} \label{eq:MissingResonances}
\end{equation}
where N is the number of resonances in the energy interval. This formula was used to estimate the number of missing resonances, by fitting the values of N and $\langle g\Gamma_{n}^{0}\rangle$, as it is presented in Fig.~\ref{fig:MissingResonances}. The result was $D_{0}$=0.66(3) eV, where the uncertainty was estimated from the statistical uncertainty due to the number of resonances considered and from calculating  $D_{0}$ in different energy ranges.

\begin{figure}[htb]
\begin{center}
\includegraphics[width=8.6cm]{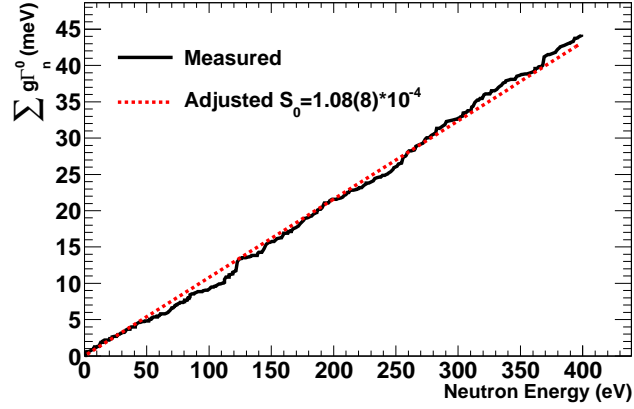} 
\caption{(Color online) Linear fit of $\sum_{\lambda}{g\Gamma_{n,\lambda}^{0}}$ as a function of the neutron energy.}
\label{fig:StrengthFunction}
\end{center}
\end{figure}

The neutron strength function for s-wave resonances $S_{0}$ can be obtained from $S_{0}=\sum_{\lambda}{g\Gamma_{n,\lambda}^{0}}/\Delta E$ and $\Delta S_{0}/S_{0}=\sqrt{2/N}$~\cite{RIPL3}, and it was calculated from the slope of the experimental cumulative sum, as it is presented in Fig.~\ref{fig:StrengthFunction}. The result was $S_{0}=1.08(8)\cdot 10^{-4}$, with an additional 3\% normalization uncertainty.

\FloatBarrier
\subsection{Analysis of the Unresolved Resonance Region}{\label{sec:URR}}

We have analyzed the energy range between 250 and 2500 eV as Unresolved Resonance Region (URR). Thus, the 250 - 400 eV energy region has been analyzed as both RRR and URR, the latter for comparison to the existing experiments and evaluations. The analysis has been performed with the SAMMY code, which contains a modified version of the FITACS code \cite{Frohner_U238,SAMMY}, which uses Hauser-Feshbach theory \cite{HauserFeshbach} with width fluctuations.

SAMMY performs the fits in the URR to the capture cross section instead of the capture yield. In the URR the shelf-shielding and multiple scattering effects are negligible, so $\sigma_{\gamma}$ was obtained directly by dividing the capture yield by the sample thickness, $\langle \sigma_{\gamma}(E_{n})\rangle =\langle Y_{\gamma}(E_{n})\rangle /n$. In the calculation of the capture yield, the background was subtracted without any smoothing procedure,  since it can not be verified if the smoothed background is at the level of the measured yield between resonances, as it can be done in the RRR. Concerning the uncertainties, all the contributions to the systematic uncertainties mentioned in Section~\ref{sec:RRR} are negligible in this energy range, with the exception of the uncertainty in the normalization. The largest contribution to the statistical uncertainties comes from the subtraction of the measured background.

The only parameters which could be fitted with the n\_TOF data were S$_{0}$ and $\langle \Gamma_{\gamma}\rangle_{0}$. The channel radius, distant level parameter $R^{\infty}_{l}$ and fission parameters are not sensitive to this measurement, and the p-wave contribution starts to be important at higher energies. In particular, according to the ENDF/B-VII.1 evaluation, the p-wave contribution to the total capture cross section is around 11\% at 2.5 keV, and a variation of 25\% in the S$_{1}$ value induces a change of only 0.5\% in the fitted value of S$_{0}$. 

The fit of S$_{0}$ and $\langle \Gamma_{\gamma}\rangle_{0}$ was performed by using the results of the statistical analysis of the RRR as prior uncertainties, and the average level spacing was fixed to the obtained value (D$_{0}$=0.66 eV). The results of the fit were S$_{0}$=1.10(4)$\cdot$10$^{-4}$ and $\langle \Gamma_{\gamma}\rangle_{0}$=42.1(20) meV, with a correlation between them of -0.23. All these uncertainties and correlations are statistical, and there is an extra systematic uncertainty of 3\% in the S$_{0}$ value due to the uncertainty in the normalization. If no prior knowledge of the parameters are assumed, compatible values of S$_{0}$ and $\langle \Gamma_{\gamma}\rangle_{0}$ are obtained, but with larger uncertainties and correlations.

\begin{figure}[htb]
\begin{center}
\includegraphics[width=8.6cm]{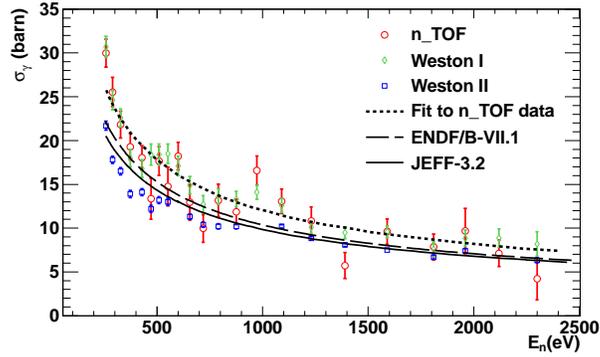} 
\caption{(Color online) Fitted n\_TOF capture data in the URR together with the two different data sets provided by Weston \textit{et al.}, in their common energy range, and with the cross sections provided by the ENDF/B-VII.1 and JEFF-3.2 evaluations.}
\label{fig:nTOFwithWeston}
\end{center}
\end{figure}

The fitted n\_TOF capture data is presented in Fig.~\ref{fig:nTOFwithWeston}, together with the only two available capture data sets at present in this energy range, tagged as ``Weston I''~\footnote{Weston \textit{et al.}, file EXFOR 12951.002, retrieved from the IAEA Nuclear Data Services website.} and ``Weston II''~\footnote{Weston \textit{et al.}, file EXFOR 12951.003, retrieved from the IAEA Nuclear Data Services website.}. Both of them have been provided by Weston \textit{et al.} (see Table~\ref{tab:differential_measurements}), in the range from 250 eV up to 92 keV and differ significantly below 1.5-2 keV. The n\_TOF data is compatible, in absolute value and shape, with the Weston I data set, whereas it is not with the Weston II data. This is an important result, since the normalization of the n\_TOF data to the available transmission experiments has been performed at low energies. On the other hand, all the present evaluations, which do not differ significantly from the two present in Fig.~\ref{fig:nTOFwithWeston}, are much closer to the Weston II data set, underestimating the $^{243}$Am capture cross section in this energy region between 7\% and 20\%.

\FloatBarrier
\subsection{Analysis at higher energies}{\label{sec:HigherEnergies}}
\FloatBarrier

The high energy limit of the n\_TOF data is 2.5 keV. However, we have used the experimental data available in EXFOR and in the literature to extend the analysis up to higher energies.

The URR ranges up to 40-42 keV in the present evaluations. As it is shown in Table~\ref{tab:differential_measurements}, there are two measurements available in EXFOR which can be used to extend the analysis of the URR up to higher energies: the one provided by Weston \textit{et al.} (above 2 keV the two datasets provided by Weston \textit{et al.} are compatible) and the one provided by Wisshak \textit{et al.}, which is 10-15\% below the Weston \textit{et al.} measurement and with a similar shape. We think that there are two reasons to prefer the normalization of Weston \textit{et al.} than the one of Wisshak \textit{et al.}. First, it is compatible with the n\_TOF results in their common energy range. Second, the Wisshak \textit{et al.} data is not compatible with the PROFIL-1 integral measurement, which is presented below. For this reason, we performed a fit to the Weston \textit{et al.} data in the 2.5-42 keV energy range by varying S$_{1}$ and $\langle \Gamma_{\gamma}\rangle_{1}$, with the values of S$_{0}$ and $\langle \Gamma_{\gamma}\rangle_{0}$ fixed to the results obtained from the n\_TOF data analysis. By doing this we are obtaining the $l=0$ parameters from the n\_TOF data, and using the Weston \textit{et al.} data to obtain the $l=1$ parameters. The results were S$_{1}$=1.65(24)$\cdot$10$^{-4}$ and $\langle \Gamma_{\gamma}\rangle_{1}$=52(34) meV, with a correlation between them of -0.82. No systematic uncertainties were included in the calculations, since their description in~\cite{Weston} is not detailed enough, and thus only the statistical uncertainties available in EXFOR were taken into account.

\begin{figure}[htb]
\begin{center}
\includegraphics[width=8.6cm]{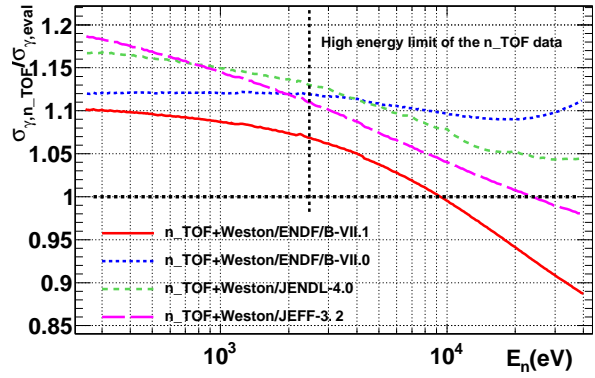} 
\caption{(Color online) Ratio between the n\_TOF fitted capture cross section and and the ones available in different evaluations. JEFF-3.2 is the same as JEFF-3.1, JENDL-3.3 and CENDL-3.1, in this energy range.}
\label{fig:Comp_URR}
\end{center}
\end{figure}

The obtained URR parameters are presented together with those obtained in other experiments and evaluations in Table~\ref{tab:Comp_URRParam}. Note that in all the cases the parameter values are at cero neutron energy, E$_{n}$=0, and in the case of the n\_TOF data the evolution of the URR parameters with E$_{n}$ is the one described in~\cite{SAMMY}. The ratio between the  results obtained in this work (from the data of n\_TOF and Weston \textit{et al.}) and the capture cross sections of different evaluations are presented in Fig.~\ref{fig:Comp_URR}. Note that the cross section obtained in this work is similar in shape to the one of the ENDF/B-VII.0 library, but 10-12\% larger.

\begin{table}[bt]
\caption{\label{tab:Comp_URRParam}
URR parameters (at E$_{n}$=0) obtained in this work compared with the ones obtained in other evaluations.
}
\begin{ruledtabular}
\begin{tabular}{lccccc}
 & D$_{0}$ & S$_{0}$ & $\langle \Gamma_{\gamma}\rangle_{0}$ & S$_{1}$ & $\langle \Gamma_{\gamma}\rangle_{1}$ \\
 & (eV) & ($\times$10$^{-4}$) & (meV) & ($\times$10$^{-4}$) & (meV) \\
\colrule
This work    & 0.66(3)   & 1.10(5)\footnotemark[1]    & 42.1(20) & 1.65(24)\footnotemark[2] & 52(34)\footnotemark[2] \\
Bellanova    & 0.62      & 0.65       &          &          &        \\
Berreth      &           &            & 42       &          &        \\
Simpson      & 0.68      & 0.96(10)   & 39       &          &        \\
Cote         &           &            & 43(3)    &          &        \\
RIPL-3~\cite{RIPL3} & 0.73(6)   & 0.98(6)    & 39(3)    &          &        \\
Mughabghab   & 0.60(6)   & 0.98(9)    & 39(1)    &          &        \\
Maslov\footnotemark[3]      & 0.57(5)   & 0.87(15)   & 43       & 2.176    & 43     \\
%Maslov       & 0.57(49) & 0.873(146) & 43       & 2.176    & 43     \\
BROND-2.2    & 0.67      & 0.93       & 39       & 2.44     & 39     \\
JENDL-4.0    & 0.44      & 0.864      & 39       & 2.687    & 39     \\
ENDF/B-VII.0 & 0.75      & 0.98       & 39       & 2.2      & 44     \\
ENDF/B-VII.1 & 0.66      & 0.98       & 39.1     & 2.6      & 69.8   \\
\end{tabular}   
\end{ruledtabular}
\footnotetext[1]{This uncertainty has been obtained by adding quadratically the statistical uncertainty ($4\cdot 10^{-4}$) to the 3\% systematic uncertainty due to the normalization.}
\footnotetext[2]{Values obtained from the n\_TOF+Weston \textit{et al.} measurements.}
\footnotetext[3]{Values adopted by the JEFF-3.2, JEFF-3.1, JENDL-3.3 and CENDL-3.1 evaluations.}
\end{table}

\begin{figure}[htb]
\begin{center}
\includegraphics[width=8.6cm]{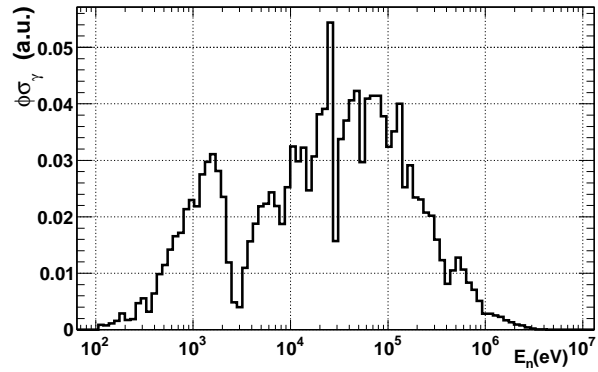} 
\caption{PROFIL-1 neutron flux multiplied by the $^{243}$Am capture cross section present in the JEFF-3.2 library.}
\label{fig:PROFIL_by_XS}
\end{center}
\end{figure}

In addition to the differential measurements of Weston \textit{et al.} and Wisshak \textit{et al.}, there are integral capture measurements in the fast energy range that can be considered for the determination of the capture cross section. One of these experiments is the PROFIL-1 irradiation experiment, where an $^{243}$Am sample were irradiated in the fast PHENIX reactor in 1974~\cite{Palmiotti_integral,ENDF/B-VII.1_benchmark}. Indeed, the changes in the ENDF/B-VII.1 evaluation with respect to ENDF/B-VII.0 were motivated by the results of this integral experiment~\cite{ENDF/B-VII.1}. The information which can be obtained from PROFIL-1 is the $^{243}$Am effective capture cross section, $\sigma_{cap}=\int{\phi (E)\sigma_{\gamma} (E)dE}$, where $\phi (E)$ is the neutron flux at the irradiated sample position.  We had access to the shape of the mentioned neutron flux, obtained from detailed Monte Carlo simulations~\cite{PROFIL_FLUX}, so we used this flux to compare the experimental values of PROFIL-1 with the ones calculated from different capture cross sections. The neutron flux used in these calculations multiplied by the $^{243}$Am capture cross section is presented in Fig.~\ref{fig:PROFIL_by_XS}, in order to show the neutron energy ranges sensitive to the PROFIL-1 integral experiment.

The references \cite{Palmiotti_integral} and \cite{ENDF/B-VII.1_benchmark} provide calculated to experimental ratios (C/E) of the mentioned effective capture cross section, $\sigma_{cap}$, each of them calculated with a different neutron data library. We did not have enough information to calculate these C/E values, but with the shape of the neutron flux we could calculate ratios between $\sigma_{cap}$ values obtained from different libraries, i.e., we could calculate ratios between different C/E values. We used the C/E value provided by \cite{ENDF/B-VII.1_benchmark} with the ENDF/B-VII.0 evaluated library to normalize our results. With this normalization, we calculated the C/E values using several evaluated data libraries, and the results obtained are presented in the second column of Table~\ref{tab:Comp_HE}. As it can be observed, they are in reasonable agreement with the values provided by the references, shown in the third and fourth columns. This indicates that the neutron flux used in this work is similar to the ones used by the references.

In a second step, we constructed several $^{243}$Am capture cross sections by taking the results obtained from the analysis of the n\_TOF+Weston data up to 42 keV (end of the URR in most of the evaluated libraries), and the energy regions above 42 keV present in the different evaluated libraries. The corresponding C/E values obtained with these cross sections (with the previously mentioned normalization) are presented in the right column of Table~\ref{tab:Comp_HE}. The experimental result has an estimated uncertainty of 5\%, so we considered that only the results obtained when using the high energy regions of the ENDF/B-VII.1, JEFF-3.2, JEFF-3.0~\cite{JEFF-3.0} and CENDL-3.1 libraries are in reasonable agreement with the PROFIL-1 integral experiment. Note that all the C/E values presented in Table~\ref{tab:Comp_HE} are below 1.

On the other hand, if the data of Wisshak \textit{et al.} are normalized to the Weston \textit{et al.} data in their common energy range (which is reasonable, since n\_TOF is compatible with Weston \textit{et al.} and because the Wisshak \textit{et al.} data is not compatible with the results of the PROFIL-1 irradiation experiment), then the capture cross sections provided by ENDF/B-VII.1 (below 100 keV) and by JEFF-3.0 (above 100 keV) are not compatible with the differential data, whereas the high energy regions ($E_{n}>42$ keV) of the JEFF-3.2 and the CENDL-3.1 libraries are compatible with them.

\begin{table}[bt]
\caption{\label{tab:Comp_HE}
C/E values of the PROFIL-1 irradiation experiment obtained with different libraries (second column), the same values provided by the references (third and fourth columns) and the C/E values obtained from the capture cross section resulting from taking the RRR and URR of this work and the part above 42 keV from the corresponding evaluated library.
}
\begin{ruledtabular}
\begin{tabular}{lcccc}
   & Library & \cite{ENDF/B-VII.1_benchmark} & \cite{Palmiotti_integral} & This work \\
 \colrule
 ENDF/B-VII.1  & 0.934 & 0.939 & & 0.939 \\
 ENDF/B-VII.0  & 0.834\footnotemark[1] & 0.834 & 0.85 & 0.889 \\
 JENDL-4.0     & 0.852 & & & 0.904 \\
 JEFF-3.2\footnotemark[2]      & 0.892 & & & 0.929 \\
 JEFF-3.0      & 0.936 & & 0.99 & 0.959 \\
 CENDL-3.1     & 0.911 & & & 0.947 \\
 ROSFOND-2010  & 0.801 & & & 0.860 \\
 ENDF/B-V.0    & 0.585 & & 0.62 & 0.730 \\
\end{tabular}   
\end{ruledtabular}
\footnotetext[1]{Value fixed to the value provided in \cite{ENDF/B-VII.1_benchmark} for normalization purposes. The rest of the values of this column were obtained from this value and the calculated C/E ratios.}
\footnotetext[2]{Same as JEFF-3.1 and JENDL-3.3.}
\end{table}

In conclusion, the $^{243}$Am capture cross section constructed from the RRR and URR obtained in this work up to 42 keV and the JEFF-3.1 or the CENDL-3.1 evaluations above 42 keV are in agreement with both the PROFIL-1 and the currently available differential capture data. In particular, this constructed cross section: (1) fits the n\_TOF data between 0.7 eV and 2.5 keV; (2) fits the mentioned ``Weston I'' data between 0.25 and 2.5 keV; (3) fits both Weston \textit{et al.} data sets between 2.5 and 42 keV; (4) fits the Wisshak \textit{et al.} data up to 250 keV, if they are normalized to the Weston \textit{et al.} data; (5) is compatible with the integral experimental results of the PROFIL-1 irradiation experiment; and (6) there is a continuous match between the URR and the high energy region, at 42 keV.

% =========================================================================================================

\FloatBarrier

\section{Conclusions}{\label{sec:Conclusions}}

The $^{243}$Am capture cross section has been measured at n\_TOF using the segmented BaF$_{2}$ Total Absorption Calorimeter (TAC), in the energy range between 0.7 and 2500 eV.

The certified mass of the $^{243}$Am sample provided by the manufacturers was not correct, and therefore we normalized the n\_TOF capture cross section to the existing transmission measurements in the neutron energy range between 3 and 50 eV. This normalization was consistent with the sample mass obtained from a high resolution $\gamma$-ray spectrometry analysis and a low resolution measurement performed with the TAC of the sample activity. In addition, this normalization is consistent with one of the only two capture measurements in the 250-2500 eV energy interval available in EXFOR. 

Due to the large flight path of the n\_TOF facility (185 m) and the statistics achieved, the results provide a better description of the resolved resonance parameters than the ones available in the current evaluated libraries, which were obtained essentially from a single transmission measurement, excluding the fission widths and the parameters of the biggest $^{243}$Am resonance at 1.35 eV. The uncertainties in the resonance parameters have been reduced, new resonances have been found and the resolved resonance region has been extended from 250 eV up to 400 eV.

The value of the resonance integral, $I_{0}=\int_{0.5eV}^{\infty}\sigma_{\gamma}(E)/EdE$, obtained in this work is significantly lower than the rest of the measured values. The strongest resonance at 1.35 eV contributes in 70-80\% to the value of $I_{0}$, and the the existence of inhomogeneities in the sample, which would affect the shelf shielding and multiple scattering corrections in this resonance , can explain this difference. The shelf shielding corrections in the rest of the resonances are very small and would not affect significantly the resulting resonance parameters.

In the unresolved resonance region, it has been found that the n\_TOF results are compatible with one of the two available capture measurements in the 0.25-2.5 keV energy range. Due to the fact that the current evaluations are closer to the other capture measurement, they underestimate the $^{243}$Am(n,$\gamma$) cross section by 7-20\% in the mentioned energy range.

We have completed the $^{243}$Am(n,$\gamma$) cross section analysis above 2.5 keV by using the data available in EXFOR and in the literature, including both differential and integral measurements. In particular, we have found an $^{243}$Am(n,$\gamma$) cross section that reproduces, under some assumptions, all the differential data sets and the PROFIL-1 integral experiment.

Taking into account the n\_TOF measurement, the $^{243}$Am(n,$\gamma$) cross section presents its larger uncertainties at thermal energies, at the strongest resonance energy of 1.35 eV, and in the fast range for reactor applications. The experimental results of Jandel \textit{et al.}~\cite{Jandel} and Hori \textit{et al.}~\cite{Hori}, which have not been published yet, could reduce further some of the mentioned uncertainties.

\begin{acknowledgments}

This work has been supported by ENRESA under the CIEMAT-ENRESA agreement on the ``Separaci\'{o}n y Transmutaci\'{o}n de Residuos Radiactivos'', the Spanish National Plan I+D+I of the Spanish Ministry of Economy and Competitiveness (project FPA2011-28770-C03-01), the European Commission 6\textsuperscript{th} Framework Programme project IP-EUROTRANS (F16W-CT-2004-516520) and the CONSOLIDER-INGENIO project CSD-2007-00042.

\end{acknowledgments}

\FloatBarrier

\FloatBarrier

% =========================================================================================================
% =========================================================================================================
%                                          BIBLIOGRAPHY
% =========================================================================================================

% The \nocite command causes allbebida entries in a bibliography to be printed out
%\nocite{*}

\bibliography{Am243_v01}

\end{document}